\documentclass[11pt]{article}
\usepackage{geometry}                
\usepackage{graphicx}                  
\graphicspath{ {/Users/jesus-pat-bloodbros/Desktop/} }
\usepackage{amssymb}
\newcommand*{\QEDA}{\hfill\ensuremath{\square}}%
\usepackage{float}
\usepackage{mathtools}
\usepackage[english]{babel}
\usepackage{amsmath}
\usepackage{epstopdf}
\usepackage{indentfirst}
\usepackage{hyperref}
\usepackage{url}
\usepackage{caption}
\numberwithin{equation}{subsection}
\title{{\bf Beating the Market with Generalized Generating Portfolios}\footnote{Acknowledgement: The author greatly acknowledges Prof. Dr. Josef Teichmann from ETH Zurich and Prof. Dr. Martin Larsson from Carnegie Mellon University for many constructive talks, fruitful discussions and helpful suggestions.}}
\date{}
\author{Patrick Mijatovic \footnote{Swiss Federal Institute of Technology (ETH Zurich), {\tt mipatric@ethz.ch}}}

\begin{document}
\pagenumbering{gobble}
\maketitle
\renewcommand{\abstractname}{Abstract}
\begin{abstract}
Stochastic portfolio theory aims at finding relative arbitrages, i.e. trading strategies which outperform the market with probability one. Functionally generated portfolios, which are deterministic functions of the market weights, are an invaluable tool in doing so. Driven by a practitioner point of view, where investment decisions are based upon consideration of various financial variables, we generalize functionally generated portfolios and allow them to depend on continuous-path semimartingales, in addition to the market weights. By means of examples we demonstrate how the inclusion of additional processes can reduce time horizons beyond which relative arbitrage is possible, boost performance of generated portfolios, and how investor preferences and specific investment views can be included in the context of stochastic portfolio theory. Striking is also the construction of a relative arbitrage opportunity which is generated by the volatility of the additional semimartingale. An in-depth empirical analysis of the performance of the proposed strategies confirms our theoretical findings and demonstrates that our portfolios represent profitable investment opportunities even in the presence of transaction costs.
\end{abstract}
\section{Introduction}
\pagenumbering{arabic}

\setcounter{page}{1}
A large part of the research in investment mathematics relies on normative assumptions on the underlying assets in the financial market. This trend was initiated by Harry Markowitz in his famous work \cite{M52}. Even though normative assumptions lead to a rich research landscape and can help in order to arrive at closed form solutions, one should try to model observed properties instead of imposing unrealistic ones, such that the discrepancy between mathematical theory and financial market realizations is kept small. This is the path Robert Fernholz went by introducing stochastic portfolio theory (SPT) in his seminal work \cite{F02}. An essential role in SPT is played by functionally generated portfolios, which represent a very robust asset allocation tool  as estimation of drifts and volatilities is absent in order to construct portfolios with controllable characteristics.

The setting of stochastic portfolio theory as introduced by Robert Fernholz only uses the market weights as inputs to the generating function. As a consequence, functionally generated portfolios are deterministic functions of the market weights. However, investment decisions are generally made by taking into consideration also other stock characteristics beside the size of a company in the market. Hence, it is desirable to have a theory in which generating functions can take as inputs additional information which could potentially take advantage of market inefficiencies. A first step into this direction was made by Strong in \cite{S13}. In this manuscript, the author proves a master equation for generating functions which in addition to the market weights depend on a process of finite variation. Additional related research was done by Schied, Speiser and Voloshchenko in \cite{SSV18}. In the named paper, the authors prove similar results to those in \cite{S13}, but in a model-free setting, by means of pathwise Ito calculus. Moreover, Ruf and Xie extend in their paper \cite{RX19} the framework of ``additive functional generation" introduced by Karatzas and Ruf in \cite{KR17}, and allow portfolio generating functions do depend on a supplementary process of finite variation. Let us also mention the recent work \cite{KK20} by Karatzas and Kim, in which the authors further expand concepts similar to those from \cite{RX19}, in a probability-free context.

The value of this work is twofold. We first generalize the master equation for generating functions which take continuous-path semimartingale arguments beyond the market weights. Second, we explicitly demonstrate how the inclusion of additional processes into generating functions can be beneficial. We come up with an example which demonstrates that this generalization can be advantageous for reduction of times beyond which relative arbitrage is possible. Furthermore, we show that it can increase the performance of generated portfolios and how one is able to implement quantitative investment preferences in the framework of stochastic portfolio theory. We also present a relative arbitrage opportunity versus the market which is generated by the volatility of the additional stock characteristics semimartingale. To the best of our knowledge these are the first such examples across SPT literature.

This paper is organized as follows. In Section 2 we summarize the main notions from stochastic portfolio theory. Specifically, we discuss stocks and portfolios in continuous time, the excess growth rate of a portfolio, the market portfolio, relative arbitrage and functional portfolio generation. In Section 3 we extend the notion of functionally generated portfolios to include continuous semimartingales as inputs, in addition to the market weights. We argue how such a generalization can be advantageous from an investor point of view. This is followed by propositions which help establish the construction of relative arbitrages in this setting. Moreover, four new examples of relative arbitrages versus the market are presented in order to illustrate our points. In Section 4 we perform a detailed empirical analysis of the proposed strategies, where we show their ability to outperform the market and SPT alternatives, not only pathwise, but also in terms of a number of performance measures. Afterwards we perform regression analysis on the three Fama French risk factors in order to understand the sources driving the returns of our portfolios. We finish this paper in Section 5, where we present a discussion of our main findings and listen the current open research problems in the field of SPT. We also suggest several new questions which were raised during the development of this work.
\section{Stochastic Portfolio Theory}
In this section we will introduce stochastic portfolio theory. We first set the stage and give the basic definitions for stocks and portfolios in continuous time. We also discuss the excess growth rate of a portfolio, which is a central quantity in SPT. Afterwards we devote a subsection to the market portfolio and its main properties. We end this section with a brief summary on functional portfolio generation.
\subsection{Stocks and Portfolios in Continuous Time}
Let $(\Omega, \mathcal{F}, \mathbb{F}, \mathsf{P})$ be a probability space with $\mathbb{F}=(\mathcal{F}_t)_{t\geq 0}$ a continuous time filtration which we assume to be right-continuous and $\mathsf{P}$-complete. Throughout this work, we shall work with this probability space. Furthermore, let $n \geq 2$ be an integer representing the number of stocks in the market. By $B=(B^1_t,...,B^n_t)_{t\geq 0}$ we denote a standard $n$-dimensional Brownian motion defined on our probability space. Hence we have $\left< B^i, B^j \right>_t = \delta^{ij}t$, $i,j=1,...,n$, for all $t\geq 0$, a.s., where $\delta^{ij}=1$ if $i=j$ and $\delta^{ij}=0$ otherwise. By $\left< X , Y\right> = (\left< X , Y\right>_t)_{t\geq 0}  $ we represent the quadratic covariation process between the stochastic processes $X=(X_t)_{t\geq 0}$ and $Y=(Y_t)_{t\geq 0}$. Compactly, we shall denote $\left< X \right> = \left<X, X \right>$. Let us write also $\mathbb{F}^B$ for the filtration generated by $B$. Note that we do not require $\mathbb{F}=\mathbb{F}^B$, but assume only that $\mathbb{F}$ contains $\mathbb{F}^B$.

We make the assumption that shares of companies are infinitely divisible. Because of this, there is no loss in generality by assuming that each firm has a single share outstanding. Thus, in this setting the stock price is equal to the total market capitalization of a company for all times. \\
\\
{\bf 2.1.1 Definition.} A {\it financial market} is a family $\mathfrak{M}=\lbrace X^1,...,X^n \rbrace$ of stocks, where $X^i = (X^i_t)_{t\geq 0}$ denotes the price process of stock $i \in \lbrace 1,...,n\rbrace$, and we postulate that it satisfies the stochastic differential equation (SDE)
\begin{equation}
\label{model}
d\log(X_t^i) := \gamma_t^idt + \sum_{j=1}^n \xi_t^{ij}dB_t^j,\,\,  t \geq 0, \,\, i=1,...,n,
\end{equation}
where the {\it growth rate} $\gamma^i$ and {\it volatilities} $\xi^{i1},...,\xi^{in}$ are $\mathbb{F}$-progressively measurable and satisfy
\begin{equation}
\int_0^T |\gamma_t^i|dt + \sum_{j=1}^n \int_0^T (\xi_t^{ij})^2 dt < \infty,\,\, T \geq 0,\,\, \text{a.s.}
\end{equation}
for each $i=1,...,n$.\\

Let us fix an $n\geq 2$ and a financial market $\mathfrak{M}=\lbrace X^1,...,X^n\rbrace$. We work with this market throughout this paper, unless stated otherwise. Moreover, we define the $\mathbb{R}^{n\times n}$-valued {\it covariance} process $\sigma = (\sigma_t^{ij})_{t\geq 0}^{1 \leq i,j \leq n}$ componentwise as
\begin{equation}
\label{covariance}
\sigma_t^{ij}:=\frac{d}{dt}\left< \log(X^i), \log(X^j)  \right>_t = \sum_{k=1}^n \xi^{ik}_t \xi^{jk}_t ,\,\,  t \geq 0,\,\, i,j=1,...,n.
\end{equation}
At this point we want to outline the importance of the covariance process $\sigma$ which describes the volatility structure of a financial market. Appropriate assumptions on the latter lead to existence of portfolios which ``outperform the market" almost surely after a finite time horizon. The two standard assumptions on the covariance process $\sigma$ under which the existence of relative arbitrage opportunities can be proven are strict nondegeneracy and sufficient intrinsic volatility. 

We say that $\mathfrak{M}$ is {\it strictly nondegenerate} if there exists an $\varepsilon >0$ such that $x \cdot \sigma_t x \geq \varepsilon x\cdot x$, for all $x\in \mathbb{R}^n$ and $t\geq 0$, almost surely, where we use the notation $x\cdot y =x^1y^1+...+x^ny^n$, $x=(1^,...,x^n)$, $y=(y^1,...,y^n)\in \mathbb{R}^n$. Strict nondegeneracy is a fairly strong condition which states that the smallest eigenvalue of the random matrix $\sigma$ should be bounded away from 0 for all times with probability one. This condition can not be expected to hold in realistic financial markets. 

The {\it sufficient intrinsic volatility} is a much weaker condition. However, we shall introduce it when we discuss the market portfolio. Next, we introduce the definitions of a {\it portfolio} and the {\it portfolio value} process.\\
\\
{\bf 2.1.2 Definition.} A {\it portfolio} $\pi = (\pi_t^1,...,\pi_t^n)_{t\geq 0}$ in the financial market $\mathfrak{M}$ is a bounded, $\mathbb{F}$-progressively measurable process which fulfils
\begin{equation}
\sum_{i=1}^n \pi_t^i =1, \,\, t\geq 0, \,\, \text{a.s.}
\end{equation}
The components of $\pi$ are called {\it portfolio weights}. We say that $\pi$ is {\it long in stock $i$ at time $t$} if $\pi_t^{i} >0$ a.s. If $\pi_t^{i} <0$ a.s., we say that {\it $\pi$ is short in stock $i$ at time $t$}. We refer to a portfolio which is long in all stocks for all times as a {\it long-only portfolio}.\\
 \\
{\bf 2.1.3 Definition.} Let $\pi$ be a portfolio in $\mathfrak{M}$. The {\it portfolio value} process $Z^{\pi} = \left( Z_t^{\pi}  \right)_{t \geq 0}$ is the process which satisfies following stochastic differential equation
\begin{equation}
\frac{dZ_t^{\pi}}{Z_t^{\pi}} := \sum_{i=1}^n \pi_t^{i} \frac{dX_t^{i}}{X_t^{i}},\,\, t > 0,\,\,  Z_0^{\pi} = Z_0,
\end{equation}
where $Z_0>0$ is the initial amount of cash invested in $\pi$.\\

We also define the {\it portfolio variance} process $\sigma^{\pi \pi} = (\sigma_t^{\pi \pi})_{t\geq 0}$ as
\begin{equation}
\sigma_t^{\pi\pi} := \sum_{i,j=1}^n \pi_t^i \sigma_t^{ij}\pi_t^j, \,\, t\geq 0.
\end{equation}
The next proposition gives the logarithmic representation of the value process of a portfolio. \\
\\
{\bf 2.1.4 Proposition.} Let $\pi$ be a portfolio in $\mathfrak{M}$. Then the portfolio value process $Z^{\pi}$ satisfies
\begin{equation}
\label{valueProcess}
d\log(Z_t^{\pi}) = \sum_{i=1}^n \pi_t^i d\log(X_t^i) + \gamma_t^{\pi, *} dt ,\,\, t \geq 0,\,\, \text{a.s.},
\end{equation}
where $\gamma^{\pi, *}=\left( \gamma_t^{\pi,*} \right)_{t\geq 0}$ is the {\it excess growth rate of a portfolio}, given by
\begin{equation}
\gamma^{\pi, *}_t :=  \frac{1}{2}\left( \sum_{i=1}^n \pi_t^i \sigma_t^{ii} - \sigma_t^{\pi \pi} \right)  ,\,\, t \geq 0.
\end{equation}
{\it Proof.} See proof of  Proposition 1.1.5 and Corollary 1.1.6 in \cite{F02}.\QEDA\\

From Proposition 2.1.4 we infer that the excess growth rate of a portfolio is a central quantity as it influences directly the portfolio value process. It will appear again when we characterize the performance of a portfolio versus the market portfolio (see Proposition 2.2.3). In the following we aim to show that the excess growth rate can be expressed conveniently by means of the {\it relative covariance process}.\\
\\
{\bf 2.1.5 Definition.} Let $\pi$ be a portfolio in $\mathfrak{M}$. We define the $\mathbb{R}^{n\times n}$-valued {\it relative covariance} process $\tau^{\pi} = (\tau_t^{\pi, ij})_{t\geq 0}^{ 1\leq i,j \leq n}$ componentwise as
\begin{equation}
\label{tau}
\tau_t^{\pi, ij} := \sigma^{ij}_t -\sigma^{i \pi}_t - \sigma^{j \pi}_t + \sigma^{\pi \pi}_t, \,\, t\geq 0, \,\, i,j=1,...,n,
\end{equation}
where the process $\sigma^{i \pi}_t$ is defined by
\begin{equation}
\sigma^{i \pi}_t := \sum_{j=1}^n \pi^{j}_t \sigma_t^{ij},\,\,\,\  t \geq 0 ,\,\,  i=1,...,n.
\end{equation}
{\bf 2.1.6 Proposition.} Let $\pi$ be a portfolio in $\mathfrak{M}$. Then $\tau^{\pi}_t$ is positive semidefinite for all $t \geq 0$, a.s. Moreover, the kernel of $\tau_t^{\pi}$ is spanned by $\pi_t$.\\
\\
{\it Proof.} See proof of Lemma 1.2.2 in \cite{F02}.\QEDA \\

The value of the relative covariance process lies in the fact that it can be used very conveniently to describe the properties of the excess growth rate. This is illustrated by the following proposition, respectively corollary. \\
\\
{\bf 2.1.7 Proposition. (Numeraire invariance property)} Let $\pi$ and $\zeta$ be any two portfolios in $\mathfrak{M}$. Then the excess growth rate fulfils
\begin{equation}
\label{numeraireInvariance}
\gamma_t^{\pi, *} = \frac{1}{2} \left( \sum_{i=1}^n \pi_t^i \tau_t^{\zeta, ii} - \sum_{i,j=1}^n \pi_t^i\tau_t^{\zeta, ij} \pi_t^j \right),
\end{equation}
for all $t\geq 0$, almost surely.\\
\\
{\it Proof.} See proof of Lemma 1.3.4 in \cite{F02}. \QEDA \\
\\
The numeraire invariance property can now be used to express the excess growth rate of a portfolio in a very compact form. \\
\\
{\bf 2.1.8 Corollary.} Let $\pi$ be a portfolio in $\mathfrak{M}$. Then for all $t\geq 0$, it holds a.s. that
\begin{equation}
\label{gamma compact}
\gamma_t^{\pi, *} = \frac{1}{2} \sum_{i=1}^n \pi_t^{i} \tau_t^{\pi, ii}.
\end{equation}
{\it Proof.} See proof of Corollary 1.3.6 in \cite{F02}. \QEDA

\subsection{The Market Portfolio}
This subsection is devoted to discuss the single most important portfolio in SPT, namely the market portfolio. We provide the definition of the market portfolio, relative arbitrage and discuss how performance of a portfolio with respect to the market can be measured. Moreover, we state the definition of market diversity and the sufficient intrinsic volatility.\\
\\
{\bf 2.2.1 Definition.} For the financial market $\mathfrak{M}$, the {\it market portfolio} $\mu = (\mu_t^1,...,\mu_t^n)_{t \geq 0}$ is defined by 
\begin{equation}
\label{marketweights}
\mu_t^{i} := \frac{X_t^{i}}{X_t^1+...X_t^n} ,\,\,  t \geq 0 ,\,\,  i=1,...,n.
\end{equation}
The components of the market portfolio $\mu^1,...,\mu^n$ are called the {\it market weights}.\\

It is clear that the market portfolio satisfies Definition 2.1.2. Furthermore, the market weights are determined by the ratio between the company's market capitalization and the market capitalization of all stocks in the market. Hence, the market weight of a stock tells us the proportion of the market that a firm constitutes in terms of capitalization. Throughout this work we shall assume that the market portfolio process is $\bigtriangleup_+^n$-valued with probability one, where
\begin{equation}
\bigtriangleup_+^n := \left\lbrace x\in (0,1)^n: \sum_{i=1}^nx^i=1 \right\rbrace.
\end{equation}
Hence, we exclude the possibility of a default of any company in $\mathfrak{M}$. It can be easily checked that the value process of the market portfolio satisfies
\begin{equation}
dZ_t^{\mu} = dX_t^1+...+dX_t^n,\,\,  t \geq 0 ,\,\,  \text{a.s.}
\end{equation}
All through this paper we shall use $\mu$ to denote the market portfolio and $Z^{\mu}$ to denote its value process. Also, we use the simpler notation $\tau$ for $\tau^{\mu}$.

The goal of SPT is to find portfolios which outperform the market portfolio over a certain time horizon with probability one. This is captured in the notion of relative arbitrage.\\
\\
{\bf 2.2.2 Definition.} Let $\pi$ and $\zeta$ portfolios in $\mathfrak{M}$ and $T \in (0, \infty)$. We say that {\it $\pi$ is a relative arbitrage opportunity versus $\zeta$ over $[0, T]$} if
\begin{equation}
\label{relarb}
\mathsf{P}(Z_T^{\pi} \geq Z_T^{\zeta}) = 1 \,\, \text{and} \,\, \mathsf{P}(Z_T^{\pi} > Z_T^{\zeta})>0
\end{equation}
hold whenever the two portfolios start with the same initial amount of cash $Z_0^{\pi}=Z_0^{\zeta}$. We shall refer to $\pi$ as a {\it strong relative arbitrage opportunity versus $\zeta$} if $\mathsf{P}(Z_T^{\pi} > Z_T^{\zeta})=1$ holds true.\\

If $\pi$ is a relative arbitrage versus $\zeta$ over $[0,T]$, then $\log(Z_T^{\pi}/Z_T^{\zeta}) \geq 0$, almost surely. Inspired by this, let us define the {\it relative return process of a portfolio $\pi$ versus $\zeta$} as $\log(Z_T^{\pi}/Z_T^{\zeta})$, for $T\geq 0$. The relative return process is a very convenient way of describing the performance of one portfolio versus another. The next proposition gives insight into the dynamics of the relative return process in the case $\zeta = \mu$.\\
\\
{\bf 2.2.3 Proposition.} Let $\pi$ be any portfolio in $\mathfrak{M}$. Then the following holds a.s., for $t\geq 0$
\begin{equation}
\label{prop323}
d\log(Z_t^{\pi}/Z_t^{\mu}) = \sum_{i=1}^n \pi_t^i d\log(\mu_t^i) + \gamma_t^{\pi, *}dt.
\end{equation}
{\it Proof.} See proof of Proposition 1.2.5 in \cite{F02}. \QEDA \\

Next, we introduce the notion of market diversity. Informally, a stock market is diverse if no single company is allowed to dominate the entire market in terms of market capitalization. This means that all the market weights are bounded away from 1 for all times. Market diversity is a very weak condition which is empirically observed and holds in any reasonable financial market model.\\
\\
{\bf 2.2.4 Definition.} We call the financial market $\mathfrak{M}$ {\it diverse} if there exists a $\delta \in (0,1)$ such that for $i=1,...,n$
\begin{equation}
\mu_t^{i} \leq 1-\delta ,\,\, t\geq 0, \,\,  \text{a.s.}
\end{equation}

Even though market diversity seems like an innocent assumption on a financial market, it has strong implications and consequences. There are examples in which it helps to establish relative arbitrage or even yield certain performance benefits (see \cite{FKK05} and Example 3.2.5).

Now, we present the definition of sufficient intrinsic volatility, which is an assumption on the volatility structure of the market and leads to the existence of relative arbitrage opportunities. \\
\\
{\bf 2.2.5 Definition.} We say that the financial market $\mathfrak{M}$ is {\it weakly sufficiently volatile} if there exists a strictly increasing continuous function $\Upsilon :[0,\infty)\rightarrow [0, \infty) $ with $\Upsilon(0)=0$ and $\Upsilon(\infty)= \infty$ such that
\begin{equation}
\infty>\frac{1}{2}\int_0^T \sum_{i=1}^n \mu_t^{i}\tau_t^{ii}dt =\int_0^T\gamma_t^{\mu, *}dt \geq \Upsilon (T),
\end{equation}
for all $T \geq 0$, a.s. Furthermore, we say that $\mathfrak{M}$ is {\it sufficiently volatile} if there exists an $\varepsilon > 0$, such that for the market excess growth rate the following holds for all $t\geq 0$ almost surely
\begin{equation}
\gamma_t^{\mu, *} \geq \varepsilon.
\end{equation}

Sufficient intrinsic volatility is a much weaker condition than strict nondegeneracy and is argued to hold in real financial markets (see \cite{FK05}). We shall see a couple of instances where sufficient intrinsic volatility leads to the existence of relative arbitrage opportunities (see Example 2.3.3, Example 3.2.5, Example 3.2.6 and Example 3.3.1).
\subsection{Functional Portfolio Generation}
In this subsection we present functional generation of portfolios and an example which shows how it can be used for construction of relative arbitrage opportunities. First, we introduce convenient notation. Afterwards, we state the definition of a functionally generated portfolio and present the key theorem of SPT, which gives the desired master equation.\\

Let $X\subset \mathbb{R}^{\ell}$, $Y\subset\mathbb{R}$ be open. By $C^2(X, Y)$, $X\subset \mathbb{R}^{\ell}, Y\subset \mathbb{R}$ we denote the space of functions $f: X \rightarrow Y$ for which
\begin{equation}
( \partial^1 )^{\beta^1}...( \partial^{\ell} )^{\beta^{\ell}}f(x)
\end{equation}
is continuous for all $\beta^1+...+\beta^{\ell}\leq 2$ and $x\in X$. Here, $\partial^i$ denotes the partial derivative with respect to the $i$-th variable. Moreover, we shall use $\partial^{ij}$ and $\partial^{i,j}$ interchangeably to denote the second partial derivative with respect to the $i$-th and $j$-th variable. We use also $\mathbb{R}_{++}$ to refer to the set of strictly positive real numbers. Furthermore, we introduce the {\it set of generating functions} $\mathfrak{G}_n$ as the set of all functions $f$ for which there exists an open neighbourhood $U$ of $\bigtriangleup_+^n$, such that $f\in C^2(U, \mathbb{R}_{++})$ and $x^i \partial^i \log(f(x))$ is bounded for all $x \in \bigtriangleup_+^n$ and $i=1,...,n$.\\
\\
{\bf 2.3.1 Definition.} Let $S : \bigtriangleup_+^n \rightarrow \mathbb{R}_{++}$ be a continuous function and $\pi$ a portfolio in $\mathfrak{M}$. We say that $\pi$ is {\it functionally generated by $S$} if there exists an $\mathbb{F}$-adapted, continuous-path, finite variation process $\Theta = (\Theta_t)_{t\geq 0}$ such that 
\begin{equation}
\label{master_mu}
\log(Z_t^{\pi}/Z_t^{\mu}) = \log(S(\mu_t) / S(\mu_0)) + \Theta_t,
\end{equation}
for all $t\geq 0$, a.s. We refer to the portfolio $\pi$ as {\it generating portfolio} and the process $\Theta$ as the {\it drift} process corresponding to the {\it generating function} $S$. \\
\\
{\bf 2.3.2 Theorem.} Let $S\in \mathfrak{G}_{n}$. Then $S$ generates the portfolio $\pi$ with weights given by
\begin{equation}
\pi^{i}_t = \mu_t^{i}(\partial^{i} \log(S(\mu_t)) + 1 - \sum_{j=1}^n
\mu_t^j \partial^j \log(S(\mu_t))),
\end{equation}
for $t \geq 0$ and $i = 1,..., n$. The drift process satisfies
\begin{align}
d \Theta_t &= - \frac{1}{2S(\mu_t)}\sum_{i, j =1}^n \partial^{ij}S(\mu_t)  \mu_t^{i} \mu_t^j  \tau_t^{ij} dt,
\end{align}
for $t\geq 0$, almost surely.\\

We call expression \eqref{master_mu} with the drift process given as in Theorem 2.3.2 {\it the master equation}. We shall prove the latter theorem in the context of a more general theory of portfolio generating functions (see the proof of Theorem 3.1.2 in the appendix). In the subsequent instance we replicate an example from \cite{FK05} in order to show how Theorem 2.3.2 can be used for construction of relative arbitrage opportunities. \\
\\
{\bf 2.3.3 Example. (Relative arbitrage in sufficiently volatile models)}  This example was originally proposed in \cite{FK05}.  Let us assume that $\mathfrak{M}$ is weakly sufficiently volatile. We start by defining the {\it generalized entropy function}
\begin{equation}
S_c(x):=c-\sum_{i=1}^n x^{i}\log(x^{i})=c+S(x), \,\, x=(x^1,...,x^n)\in \bigtriangleup_+^n,
\end{equation}
for $c>0$, and $S(x)=-\sum_{i=1}^n x^{i}\log(x^{i})$ is the {\it standard entropy function}. Since $S_c \in \mathfrak{G}_n$, we can take advantage of Theorem 2.3.2. The function $S_c$ generates the portfolio
\begin{equation}
\label{entropyPortfolio}
\pi_t^{i} = \frac{c\mu_t^{i}-\mu_t^{i}\log(\mu_t^{i})}{c -\sum_{j=1}^n\mu_t^j\log(\mu_t^j)} ,\,\,  t \geq 0 ,\,\,  i=1,...,n,
\end{equation}
to which we refer as the {\it entropy weighted portfolio}. One can also check that the drift process is given by
\begin{equation}
\label{entropyDrift}
d\Theta_t = \frac{1}{2}\sum_{i=1}^n \frac{\tau_t^{ii}\mu_t^{i}}{S_c(\mu_t)}dt,\,\,  t \geq 0 ,\,\,  \text{a.s.}
\end{equation}
The claim is that the portfolio \eqref{entropyPortfolio} does the job, i.e. it is a relative arbitrage opportunity versus the market portfolio. In order to show this, we will need
\begin{equation}
c\leq S_c(\mu_t) \leq c+\log(n),\,\,  t\geq 0 ,\,\, \text{a.s.},
\end{equation}
which follows from $0\leq S(x) \leq \log(n)$, $x\in \bigtriangleup_+^n$, and $\mathsf{P}(\mu_t\in \bigtriangleup_+^n)=1$, $t\geq 0$. The master equation yields
\begin{align}
\label{master eq. entropy}
\log(Z_T^{\pi}/Z_T^{\mu})&= \log(S_c(\mu_T)/S_c(\mu_0)) +\int_0^T \frac{\gamma_t^{\mu, *}}{S_c(\mu_t)}dt\nonumber \\& \geq \log\left( \frac{c}{c+S(\mu_0)} \right) + \frac{\Upsilon(T)}{c+\log(n)} ,\,\, \text{a.s.},
\end{align}
for $T\geq 0$, where we have used the bounds on $S_c$ and the assumption that the market is weakly sufficiently volatile. The right hand side of \eqref{master eq. entropy} is strictly positive if
\begin{equation}
\label{lower bound entropy time}
T > T^*(c):=\Upsilon^{-1}\left((c+\log(n))\log\left(1+\frac{S(\mu_0)}{c} \right)\right),
\end{equation}
which is well defined since $\Upsilon$ is strictly increasing and continuous, and hence invertible. Thus, the entropy weighted portfolio \eqref{entropyPortfolio} is a strong relative arbitrage opportunity versus market over all time horizons $[0,T]$ with $T>T^*(c)$. Moreover, from
\begin{equation}
T^* := \lim_{c\rightarrow \infty} T^*(c) = \Upsilon^{-1}(S(\mu_0)),
\end{equation}
we can conclude that for each $T>T^*$ we can choose a $c$ sufficiently large, such that $\pi$ outperforms the market over $[0, T]$.
\QEDA 
\section{Generalized Generating Portfolios}
In this section we present the main results of this paper. We first introduce generalized generating functions and state the generalized master equation. Afterwards we prove a couple of propositions which indicate what functions are appropriate in order to obtain benefits in this generalized framework. We then apply these results in order to construct four novel trading strategies which outperform the market almost surely.
\subsection{The Generalized Master Equation}
We first adapt the definition of generating portfolios such that the corresponding generating functions take as inputs continuous semimartingales, beyond the market weights. We impose that the additional semimartingale is $\mathbb{F}$-progressively measurable, has continuous paths almost surely and is not equal to a deterministic function of the market weights. If this is the case, this additional process is called the {\it stock characteristics} process. Throughout this paper we denote by $K$ a subset of $\mathbb{R}^k$, for some $k\in \mathbb{N}$, and we make additional specifications on $K$ only when needed.\\
\\
{\bf 3.1.1 Definition.} Let $P$ be a stock characteristics process which is valued in $K$, let $S : \bigtriangleup_+^n \times K \rightarrow \mathbb{R}_{++}$ be a continuous function and $\pi $ a portfolio in $\mathfrak{M}$. We say that $S$ {\it generates the portfolio $\pi$ with stock characteristics $P$} if there exists an $\mathbb{F}$-adapted, continuous-path, finite variation process $\Theta$, such that the following holds
\begin{equation}
\log(Z^{\pi}_t /Z^{\mu}_t) = \log(S(\mu_t, P_t)/S(\mu_0, P_0)) - \sum_{i=1}^{k}\int_0^t  \partial ^{n+i} \log (S(\mu_s, P_s)) dP_s^{i} + \Theta_t,
\end{equation} 
for all $t\geq 0$, almost surely. We shall refer to such a portfolio as a {\it generalized generating portfolio} and the process $\Theta = (\Theta_t)_{t\geq 0}$ is called the {\it drift} process corresponding to the {\it generating function} $S$.\\

Comparing Definition 3.1.1 with Definition 2.3.1 it is evident that the generalized version has an additional integral. If the stock characteristics process $P$ is a finite variation process, then the integral is a Lebesgue-Stieltjes integral which is also of finite variation. In this case we define the {\it extended drift} process $\widetilde{\Theta}$ by
\begin{equation}
\widetilde{\Theta}_t := \Theta_t - \sum_{i=1}^k \int_0^t \partial^{n+i} \log(S(\mu_s, P_s))dP_s^{i},
\end{equation}
for $t\geq 0$. Hence, the formal structure of Definition 2.3.1 remains for finite variation stock characteristics processes. If $P$ is a semimartingale with a local martingale part which has infinite first variation, then the integral is a stochastic integral.

We further extend the set of generating functions introduced in the previous section. Let the stock characteristics process be $K$-valued. By $\mathfrak{G}_n^K$ we denote the space of all functions $f$ for which there exists an open neighbourhood $U$ of $\bigtriangleup_+^n \times K$ such that $f\in C^2(U, \mathbb{R}_{++})$, and for $i=1,...,n$, the quantity $x^i\partial^i \log(f(x,y))$ is bounded for all $(x,y) \in \bigtriangleup^n_+ \times K$.\\
\\
{\bf 3.1.2 Theorem.} Assume that $P$ is a stock characteristics process which is valued in $K$ and let $S\in \mathfrak{G}_n^K$. Then $S$ generates the portfolio $\pi$ with weights given by
\begin{equation}
\label{portfolio weights}
\pi^{i}_t = \mu_t^{i}(\partial^{i} \log(S(\mu_t, P_t)) + 1 - \sum_{j=1}^n
\mu_t^j \partial^j \log(S(\mu_t, P_t))),
\end{equation}
for $t \geq 0$ and $i = 1,..., n$. The drift process satisfies
\begin{align}
\label{drift process}
d \Theta_t &= - \frac{1}{2S(\mu_t, P_t)}\sum_{i, j =1}^n\partial^{ij}S(\mu_t, P_t) \mu_t^{i}\mu_t^j  \tau_t^{ij}dt - \frac{1}{2}\sum_{i, j=1}^{k} \partial^{n+i,n+j} \log(S(\mu_t, P_t)) d\left< P^{i}, P^j\right>_t\nonumber  \\&-\sum_{i=1}^n\sum_{j=1}^{k} \partial^{i,n+j} \log(S(\mu_t, P_t)) d\left< \mu^{i}, P^j\right>_t,
\end{align}
for all $t\geq 0$, almost surely.\\

We provide a proof of Theorem 3.1.2 in the appendix. Expression (3.1.1) along with the drift process given by Theorem 3.1.2 is referred to as the {\it generalized master equation}.\\
\\
{\bf Remarks.} From the generalized master equation one is able to obtain many advantages compared to the classical one. We already have two benefits:
\begin{itemize}
\item For a fixed time horizon $[0,T]$, for some $T>0$, an appropriately chosen generating function $S$ and stock characteristics process $P$ can increase the outperfomance relative to the market via the additional terms in the drift process.
\item For a fixed goal of outperformance, i.e. $\log(Z_T^{\pi} / Z_T^{\mu}) \geq c$, for some $c>0$, by an appropriate choice of $S$ and $P$ one can reduce the time $T$ by which this goal is achieved.
\end{itemize}

The generalized master equation is an ideal framework for comparing the performance of functionally generated portfolios to the market portfolio. However, comparison to other benchmark portfolios is possible as well. Usually these benchmark portfolios are very simple and do not depend on additional stock characteristics beside the market weights. The next proposition gives insight how this can be done.\\
\\
{\bf 3.1.3 Proposition.} Let the stock characteristics process $P$ be valued in $K$ and assume that $S^{\zeta}\in \mathfrak{G}_n$ and $S^{\pi}\in \mathfrak{G}_n^K$. Let $\zeta, \pi$ be the portfolios generated by $S^{\zeta}, S^{\pi}$ respectively. Then the following holds true for $t\geq 0$, almost surely
\begin{equation}
\label{eq4120}
d\log(Z^{\pi}_t/Z_t^{\zeta})= d\log(S^{\pi}(\mu_t, P_t)/S^{\zeta}(\mu_t)) -\sum_{i=1}^k\partial^{n+i}\log(S^{\pi}(\mu_t, P_t))dP_t^{i} + d\Theta_t^{\pi} - d\Theta_t^{\zeta},
\end{equation}
where $\Theta^{x}$ denotes the drift process of portfolio $x \in \lbrace \pi, \zeta \rbrace$. \\
\\
{\it Proof.} Equation \eqref{eq4120} follows from $d\log(Z_T^{\pi}/Z_T^{\zeta}) =d\log(Z_T^{\pi}/Z_T^{\mu}) - d\log(Z_T^{\zeta}/Z_T^{\mu})$ and Definition 3.1.1. \QEDA \\
\\
{\bf 3.1.4 Example. (Equally weighted portfolio as benchmark)} We use here the notation from Proposition 3.1.3. The {\it equally weighted portfolio (EWP)} is a prominent benchmark in the financial industry. It is defined as
\begin{equation}
\zeta_t^{i} := \frac{1}{n},
\end{equation}
for $i=1,..., n$ and all $t\geq 0$. The equally weighted portfolio is functionally generated by
\begin{equation}
S(x)=(x^1 \cdot ... \cdot x^n)^{1/n}, \,\, x=(x^1,...,x^n)\in\bigtriangleup_+^n.
\end{equation}
Furthermore, the drift process corresponding to $S$ satisfies for $t\geq 0$, a.s.
\begin{equation}
d\Theta_t^{\zeta}=\frac{1}{2}\left( \frac{1}{n}\sum_{i=1}^n\tau_t^{ii}-\frac{1}{n^2}\sum_{i,j=1}^n \tau_t^{ij}     \right)dt=\gamma_t^{\zeta, *}dt,
\end{equation}
where we have used the numeraire invariance property in the last equality. Since the EWP is a long-only portfolio, we have $\gamma_t^{\zeta, *}>0$ for all $t\geq0$ a.s. (see Proposition 1.3.7 in \cite{F02}). This implies then for all $T>0$, that $\Theta_T=\int_0^T\gamma_t^{\zeta, *} dt>0$, a.s., which explains the surprisingly good performance of the EWP.

Let now the stock characteristics process be valued in $K$, and choose an $S^{\pi}\in \mathfrak{G}_n^K$. Proposition 3.1.3 states that for the value process of the portfolio $\pi$ generated by $S^{\pi}$ and the portfolio $\zeta$ the following holds true
\begin{align}
\label{EWP}
\log(Z_T^{\pi}/Z_T^{\zeta})&=\log\left( \frac{S_T^{\pi}(\mu_0^1\cdot ...\cdot \mu_0^n)^{1/n}}{S_0^{\pi}(\mu_T^1\cdot...\cdot\mu_T^n)^{1/n}}   \right) - \sum_{i=1}^k\int_0^T \partial^{n+i}\log(S(\mu_t, P_t))dP_t^{i}\nonumber \\ &+\Theta_T^{\pi}-\int_0^T\gamma_t^{\zeta, *}dt,
\end{align}
for all $T\geq 0$, almost surely. Note the strictly negative contribution of the last term in \eqref{EWP}. It is for this reason, why it is a difficult task to find portfolios in practice which beat the EWP. Note also that \eqref{EWP} is well defined due to the standing assumption that no company in $\mathfrak{M}$ defaults. \QEDA
 \subsection{Applications of the Generalized Master Equation}
In this subsection we want to give examples and applications of the generalized master equation. We are mainly interested to motivate why the inclusion of additional stock characteristics processes into generating functions can be beneficial. Before doing so, we prove two lemmas which give insight into which functions are appropriate for our discussion.\\
\\
{\bf 3.2.1 Definition.} Let $S:U_{\bigtriangleup_+^n}\times U_K \rightarrow \mathbb{R}_{++}$ be a continuous function, where $U_{\bigtriangleup_+^n}$ is an open neighbourhood of $\bigtriangleup_+^n$ and $U_K$ is an open neighbourhood of $K$. We say that $S$ is {\it multiplicative} if there exist continuous functions $f : U_{\bigtriangleup_+^n} \rightarrow \mathbb{R}_{++}$ and $g : U_K \rightarrow \mathbb{R}_{++}$ such that $S(x, y) = f(x)g(y)$ for all $(x,y)\in U_{\bigtriangleup_+^n}\times U_K$.\\
\\
{\bf 3.2.2 Lemma.} Assume that the stock characteristics process $P$ is $K$-valued almost surely. Let $S$ be a multiplicative function with $S\in \mathfrak{G}_n^K$. Then neither the portfolio weights $\pi$ generated by $S$, nor the return of $\pi$ relative to the market depend on $P$.\\
\\
{\it Proof.} Let the stock characteristics process $P$ be $K$-valued almost surely and let $S$ be multiplicative and $S \in \mathfrak{G}_n^K$. By definition there exist functions $f\in C^2(U_{\bigtriangleup_+^n}, \mathbb{R}_{++})$ and $g\in C^2(U_K, \mathbb{R}_{++})$ with $U_{\bigtriangleup_+^n} \supset \bigtriangleup_+^n$ and $U_K \supset K$ open such that $S(x,y)=f(x)g(y)$, for all $(x,y) \in \bigtriangleup_+^n \times K$. Using this it is easy to see that the identities
\begin{equation}
\label{first one}
\partial^{i}\log(S(\mu_t, P_t)) = \partial^{i} \log(f(\mu_t)),\,\, i=1,...,n,
\end{equation}
\begin{equation}
\label{second one}
\frac{\partial^{ij}S(\mu_t, P_t)}{S(\mu_t)}=\frac{\partial^{ij}f(\mu_t, P_t)}{f(\mu_t)},\,\, i,j=1,...,n,
\end{equation}
\begin{equation}
\label{third one}
\partial^{n+i} \log(S(\mu_t, P_t)) = \partial^{i} \log(g(P_t)),\,\, i=1,...,k,
\end{equation}
\begin{equation}
\label{forth one}
\partial^{i,n+j} \log(S(\mu_t, P_t)) =0,\,\, i=1,...,n,\,\, j=1,...,k,
\end{equation}
hold true for $t\geq 0$, a.s. Since $S$ is assumed to be a member of $\mathfrak{G}_n^K$, we can apply Theorem 3.1.2. Using identity \eqref{first one} we can deduce that the portfolio weights generated by $S$ are given by
\begin{equation}
\label{independent weights}
\pi_t^{i}=\mu_t^{i}(\partial^{i}\log(f(\mu_t)) + 1 -\sum_{j=1}^n \mu_t^{j}\log(f(\mu_t))),\,\, t \geq 0,
\end{equation}
for $i=1,...,n$, which is independent of $P$. The generalized master equation reads
\begin{align}
\label{independent rel.return}
d\log(Z_t^{\pi}/Z_t^{\mu}) &= d\log(f(\mu_t)) -\frac{1}{2f(\mu_t)}\sum_{i,j=1}^n\partial^{ij}f(\mu_t)\tau_t^{ij}\mu_t^{i}\mu_t^{j}dt \nonumber \\& +d\log(g(P_t)) -\sum_{i=1}^k \partial^{i}\log(g(P_t))dP_t^i -\frac{1}{2} \sum_{i, j=1}^k \partial^{ij}\log(g(P_t))d\left<P^{i},P^{j}\right>_t \nonumber \\&= d\log(f(\mu_t)) -\frac{1}{2f(\mu_t)}\sum_{i,j=1}^n\partial^{ij}f(\mu_t)\tau_t^{ij}\mu_t^{i}\mu_t^{j}dt,\,\, t\geq 0,\,\, \text{a.s.},
\end{align}
where the identities \eqref{second one}, \eqref{third one} and \eqref{forth one} helped us in the first equality. In order to arrive at the second equality, we have taken advantage of Ito's lemma applied on $\log(g(P_t))$. This gives the desired claim. \QEDA \\
\\
{\bf 3.2.3 Corollary.} Let the stock characteristics process $P$ be valued in $K$ and $S\in \mathfrak{G}_n^K$. Choose any $F\in C^2(U_K, \mathbb{R}_{++})$, where $U_K$ is an open neighbourhood of $K$. Let $\pi$ and $\zeta$ denote the portfolios generated by $S$ and $S\cdot F$, respectively. It holds true that
\begin{equation}
\pi_t^{i}=\zeta_t^i, \,\, t\geq 0, \,\, i=1,...,n, ,\,\, \text{a.s.},
\end{equation}
and hence also $Z_t^{\pi} =Z_t^{\zeta}, \,\, t\geq 0, \,\, \text{a.s.}$\\
\\
{\bf Remarks.} Once we have proven \eqref{independent weights}, namely that for multiplicative generating functions the portfolio weights do not depend on the stock characteristics process, an intuitive reasoning suggests that the relative returns versus the market can not depend on the additional characteristics either. This is then confirmed also rigorously by \eqref{independent rel.return}.\\

Lemma 3.2.2 and Corollary 3.2.3 play an important role in the discussion to come since they restrict the type of functions for which we can obtain the benefits of including additional processes as arguments of generating functions. We conclude that only functions which are non-multiplicative are able to give additional performance benefits compared to the classical SPT approach where portfolios are constructed solely based on market weights.

In general, relative arbitrages versus the market can be generated by functions for which the drift process is increasing. Propostion 3.1.15 and Proposition 3.4.2 in \cite{F02} state that the drift process is increasing for concave generating functions. Hence, these functions are good candidates to generate portfolios which outpeform the market. However, this statement is only valid in the absence of stock characteristics processes. The next lemma generalizes Proposition 3.4.2 from \cite{F02} in the case of generating functions which additionally take increasing or decreasing stock characteristics processes as inputs. For a function $S\in\mathfrak{G}_n^K$, we say that $S$ is {\it concave} on $\bigtriangleup_+^n$ if for every fixed $y\in K$ we have that $S(\lambda x_1 + (1-\lambda)x_2, y)\geq \lambda S(x_1, y) + (1-\lambda)S(x_2,y)$, for all $\lambda \in (0,1)$ and any $x_1, x_2 \in \bigtriangleup_{+}^n$. Moreover, we say that $S$ is increasing (decreasing) on $K$ if for every fixed $x\in\bigtriangleup_+^n$ it holds that $\partial^{n+i}S(x,y)\geq 0$ ($\partial^{n+i}S(x,y)\leq 0$), for all $i=1,...,k$ and $y\in K$.
\\
\\
{\bf 3.2.4 Lemma.} Assume that the stock characteristics process $P$ is either increasing or decreasing and $K$-valued. Let $S\in \mathfrak{G}_n^K$ be a non-multiplicative function which is concave on $\bigtriangleup_+^n$ and has the opposite monotonicity of $P$ on $K$. Then the extended drift process $\widetilde{\Theta}$ is increasing almost surely.\\
\\
{\it Proof.} Assume that $P$ is an increasing $K$-valued stock characteristics process and let $S \in \mathfrak{G}_n^K$ be non-multiplicative, concave on $\bigtriangleup_+^n$ and decreasing on $K$. Recall that for a finite variation stock characteristics process $P$ we define the extended drift process by 
\begin{equation}
\label{extendedDrift}
\widetilde{\Theta}_T = -\frac{1}{2}\sum_{i,j=1}^n\int_0^T \frac{\partial^{ij}S(\mu_t, P_t)}{S(\mu_t, P_t)}\mu_t^i \mu_t^j \tau_t^{ij}dt - \sum_{i=1}^k\int_0^t  \partial^{n+i}\log(S(\mu_t, P_t))dP_t^{i},\,\, T \geq 0.
\end{equation}
We will show that both integrals in \eqref{extendedDrift} are increasing in $T$. Regarding the first integral in \eqref{extendedDrift}, we follow the proof of Proposition 3.1.15 in \cite{F02}. Let us denote by $HS_y(x):=(\partial^{ij}S(x,y))_{1\leq i,j \leq n}$ the Hessian matrix of $S$ with respect to the first $n$ variables of $S$ evaluated at $x\in \bigtriangleup_+^n$ for a fixed $y\in K$. Let now $t\geq 0$ be arbitrary and observe that $HS_{P_t}(\mu_t)$ is diagonalizable with eigenvalues $\lambda^1,...,\lambda^n$ and corresponding normalized eigenvectors $e^1,...,e^n$. Hence we may write
\begin{equation}
\left(HS_y(x)\right)^{ij} =\partial^{ij}S(\mu_t, P_t)= \sum_{\ell = 1}^n \lambda^{\ell} e^{\ell i}e^{\ell j},
\end{equation}
where $e^{\ell i}$ denotes the $i$-th component of the eigenvector $e^{\ell}$, for $i,\ell = 1,...,n$. Then it follows that
\begin{equation}
\sum_{i,j=1}^n \partial^{ij} S(\mu_t, P_t) \mu_t^i \mu_t^j \tau_t^{ij} = \sum_{\ell =1}^n \lambda^{\ell} \sum_{i,j=1}^n e^{\ell i}e^{\ell j}\mu_t^i \mu_t^j \tau_t^{ij} \leq 0,
\end{equation}
where the inequality follows from the fact that $\lambda^1,...,\lambda^n \leq 0$, since $S$ is concave on $\bigtriangleup_+^n$, and from Proposition 2.1.6 which states that the $\mathbb{R}^{n\times n}$-valued relative covariance process $\tau$ is positive semidefinite for all times. Hence, since $t\geq 0$ was arbitrary, we get that
\begin{equation}
-\frac{1}{2}\sum_{i,j=1}^n\int_0^T  \frac{\partial^{ij}S(\mu_t, P_t)}{S(\mu_t, P_t)}\mu_t^{i}\mu_t^j \tau_t^{ij}dt \geq 0
\end{equation}
and is increasing as a function of $T$, almost surely. It remains to prove that the second integral in \eqref{extendedDrift} is increasing as well. Observe that for any $t\geq 0$ and any $i=1,...,k$, $\partial^{n+i}S(\mu_t, P_t) \leq 0$, almost surely, since $S$ is assumed to be decreasing on $K$. This however means that
\begin{equation}
-\sum_{i=1}^k\partial ^{n+i} \log(S(\mu_t, P_t)) =\sum_{i=1}^k \frac{\partial^{n+i}S(\mu_t, P_t)}{S(\mu_t, P_t)} \geq 0.
\end{equation}
Combining this with the fact that $P$ is increasing and that $t \geq 0$ was arbitrary yields that
\begin{equation}
-\sum_{i=1}^k \int_0^T \partial^{n+i} \log(S(\mu_t, P_t))dP_t^i \geq 0,
\end{equation} 
and is increasing as a function of $T$, almost surely. Hence, the extended drift process $\widetilde{\Theta}$ is positive and increasing as wished. The proof for the case of $P$ being decreasing and $S$ increasing on $K$ is analogous. \QEDA \\

In the sequel we examine two examples to illustrate the benefits arising from the generalized master equation. 

Recall Example 2.3.3, where it is shown that in a market which is weakly sufficiently volatile, relative arbitrage versus the market exists for sufficiently long time horizons. In the example below we show that the lower bound on the time horizon \eqref{lower bound entropy time} can be shortened using the generalized master equation. We shall demonstrate how the inclusion of a single stock characteristics process can already generate a strictly positive contribution to the relative return versus the market. This in turn leads to the reduction of the minimal time horizon beyond which relative arbitrage is possible.\\
\\
{\bf 3.2.5 Example. (Reduction of relative arbitrage times)} Let us assume the market $\mathfrak{M}$ is diverse and weakly sufficiently volatile. Recall from Definition 2.2.5 that the latter proposition holds if there exists a strictly increasing continuous function $\Upsilon : [0, \infty)\rightarrow [0,\infty)$, such that for the market excess growth rate $\gamma^{\mu, *}$ the following holds almost surely
\[  \int_0^T\gamma_t^{\mu,*}dt \geq \Upsilon(T), \,\, T\geq 0.  \]

We consider time as our stock characteristics process, i.e. in terms of the notation of Definition 3.1.1 we have $P_t=t$ and $K=[0, \infty)$. 

Since the market is diverse, it follows from Proposition 2.3.2 in \cite{F02} that there exists an $\varepsilon>0$ such that
\begin{equation}
\label{entropybound}
S(\mu_t)\geq\varepsilon,\,\, t \geq 0,\,\, \text{a.s.},
\end{equation}
where $S(\mu_t):=-\sum_{i=1}^n \mu_t^{i}\log(\mu_t^{i})$ is the {\it market entropy} at time $t$, for $t\geq 0$.
Consider now the generating function $\widetilde{S}_c$ defined by
\begin{equation}
\widetilde{S}_c(x, y): = c -\sum_{i=1}^nx^{i}\log(x^{i}) -\varepsilon\tanh(y) = S_c(x) -\varepsilon\tanh(y),
\end{equation}
for $x=(x^1,...,x^n)\in\bigtriangleup_+^n$ and $y\in [0, \infty)$. Here, $S_c(x)=c+S(x)$ denotes the generalized entropy function from Example 2.3.3, $c>0$ is a constant and $\varepsilon$ is the market entropy lower bound from \eqref{entropybound}. It is evident that $\widetilde{S}_c \in \mathfrak{G}_n^{[0, \infty)}$. By this, we are allowed to take advantage of Theorem 3.1.2. The function $\widetilde{S}_c$ generates the portfolio
\begin{equation}
\label{portfolio reduction}
\pi_t^{i} = \frac{\mu_t^{i}(c-\varepsilon\tanh(t) -\log(\mu_t^{i}))}{c-\varepsilon\tanh(t) -\sum_{j=1}^n\mu_t^j\log(\mu_t^j)},\,\, t \geq 0,\,\, i=1,...,n.
\end{equation}
The fact that our stock characteristics process is of finite variation significantly simplifies the drift term \eqref{drift process}, as only the first sum remains. The drift process is determined by
\begin{equation}
\label{drift reduction}
d\Theta_t=\frac{1}{2}\sum_{i=1}^n \frac{1}{\widetilde{S}_c(\mu_t, t)}\tau_t^{ii}\mu_t^{i}dt,\,\, t \geq 0, \,\, \text{a.s.}
\end{equation}
The following bound on $\widetilde{S}_c$ will prove to be useful
\begin{equation}
\label{entropy estimate}
c \leq \widetilde{S}_c(\mu_t, t) \leq c+\log(n),\,\,  t \geq 0,\,\, \text{a.s.}
\end{equation}
This follows from $-\varepsilon \leq -\varepsilon\tanh(y) \leq 0$, $y\in [0, \infty)$, and $\varepsilon\leq S(x)\leq \log(n)$, $x\in\bigtriangleup_+^n$. Hence, the drift process satisfies a.s.
\begin{equation}
\label{drift estimate}
\Theta_T \geq \frac{1}{c+\log(n)}\int_0^T \gamma_t^{\mu,*}dt \geq \frac{1}{c+\log(n)} \Upsilon(T), \,\, T\geq 0,
\end{equation} 
where in the last inequality we have used the weak sufficient volatility assumption. The master equation reads
\begin{equation}
\label{master reduction}
\log(Z_T^{\pi}/Z_T^{\mu}) = \log(\widetilde{S}_c(\mu_T, T)/\widetilde{S}_c(\mu_0,0))+ \Theta_T +\varepsilon \int_0^{T}\frac{1-\tanh^2(t)}{\widetilde{S}_c(\mu_t, t)}dt,
\end{equation}
for $T \geq 0$, a.s. In order to arrive at \eqref{master reduction}, the identity $d(\tanh(x))/dx = 1-\tanh^2(x)$, $x\in\mathbb{R}$, has helped us. The integral on the right hand side of \eqref{master reduction} is a Lebesgue Stieltjes integral and can be estimated thanks to \eqref{entropy estimate} as follows
\begin{equation}
\label{integral estimate}
\int_0^T\frac{(1-\tanh^2(t))}{\widetilde{S}_c(\mu_t, t)}d t\geq  \frac{\tanh(T)}{c+\log(n)},\,\, T \geq 0,\,\,  \text{a.s.},
\end{equation}
where we have used $\tanh(0)=0$. Using \eqref{entropy estimate}, \eqref{drift estimate} and \eqref{integral estimate} in the master equation \eqref{master reduction}, we conclude that the following holds almost surely for the relative return of the portfolio $\pi$ versus the market
\begin{align}
\label{master reduction estimate}
\log(Z_T^{\pi}/Z_T^{\mu}) &\geq \log\left( \frac{c}{c+S(\mu_0)} \right) + \ \frac{\Upsilon(T)}{c+\log(n)} +\frac{\varepsilon}{c+\log(n)}\tanh(T),
\end{align}
for all $T\geq 0$. Let us define the function
\begin{equation}
\widetilde{\Upsilon}(T)=\Upsilon(T) +\varepsilon\tanh(T), \,\, T \geq 0,
\end{equation}
which is strictly increasing and continuous, and hence it possesses an inverse. From \eqref{master reduction estimate} it follows then that after a time $T$ which satisfies
\begin{equation}
T >\widetilde{T}(c):= \widetilde{\Upsilon}^{-1}\left((c+\log(n))\log\left(1+\frac{S(\mu_0)}{c} \right) \right),
\end{equation}
the portfolio \eqref{portfolio reduction} will outperform the market almost surely, for all $c>0$. Similar to the final argument in Example 2.3.3, we can deduce that for all $T> \widetilde{T} := \widetilde{\Upsilon}^{-1}(S(\mu_0))$, we can choose a $c>0$ large enough, such that $\pi$ ``beats" the market over $[0, T]$.

Under the assumption that the $\Upsilon$ from Example 2.3.3 and from this example coincide, let us compare the times $\widetilde{T}(c), \widetilde{T}$ with the times $T^*(c), T^*$ from Example 2.3.3. Recall that $T^*(c)$ and $T^*$ are given by
\begin{equation*}
T^*(c)=\Upsilon^{-1}\left((c+\log(n))\log\left(1+\frac{S(\mu_0)}{c} \right)  \right),
\end{equation*}
\begin{equation*}
T^*=\Upsilon^{-1}(S(\mu_0)).
\end{equation*}
Since $\widetilde{\Upsilon}(T)>\Upsilon(T) $, it follows that $\widetilde{\Upsilon}^{-1}(T) < \Upsilon^{-1}(T)$, for all $T > 0$.
Hence, we conclude that $\widetilde{T}(c) < T^*(c)$, for all $c>0$, and especially $\widetilde{T} < T^*$. Therefore, considering stock characteristics processes and the generalized master equation can lead to reduction of time horizons beyond which a certain goal of outperformance is desired. \QEDA \\

In the next instance we give an example of a generalized generating portfolio which outperforms the entropy weighted portfolio from Example 2.3.3.\\
\\
{\bf 3.2.6 Example. (Beating the entropy weighted portfolio)} Suppose that the market $\mathfrak{M}$ is weakly sufficiently volatile and let $\zeta$ denote the entropy weighted portfolio and $S_c$ the generalized entropy function from Example 2.3.3. Moreover, $\Upsilon:[0, \infty)\rightarrow [0, \infty)$ is the strictly increasing continuous function which represents the lower bound on the cumulative excess growth rate of the market. For a fixed $\alpha\in(0, 1/2)$, we define $\widetilde{S}_c$ by
\begin{equation}
\widetilde{S}_c(x,y) := c-\sum_{i=1}^n x^i\log(x^i) +\alpha c(\tanh(-y)-1), \,\, x=(x^1,...,x^n) \in \bigtriangleup_+^n ,\,\, y\in [0, \infty).
\end{equation}
The latter generates the portfolio
\begin{equation}
\label{entropy boost portfolio}
\pi_t^{i} = \frac{\mu_t^{i}(c+\alpha c(\tanh(-t)-1) -\log(\mu_t^{i}))}{c+\alpha c(\tanh(-t)-1) -\sum_{j=1}^n\mu_t^j\log(\mu_t^j)},\,\, t \geq 0,\,\, i=1,...,n.
\end{equation}
Moreover, the drift process reads
\begin{equation}
\label{drift comparison}
d\Theta_t=\frac{1}{2}\sum_{i=1}^n \frac{1}{\widetilde{S}_c(\mu_t, t)}\tau_t^{ii}\mu_t^{i}dt,\,\, t \geq 0, \,\, \text{a.s.}
\end{equation}
Let us also comment that $\widetilde{S}_c$ admits the following bounds
\begin{equation}
c-2\alpha c\leq \widetilde{S}_c(\mu_t, t ) \leq c+\log(n) - \alpha c, \,\, t\geq 0,\,\, \text{a.s.}
\end{equation}
We will show that $\pi$ outperforms $\zeta$ over a sufficiently long time horizon. From the definition of $S_c$ and $\widetilde{S}_c$ and the identity $-\tanh(-x)=\tanh(x)$, $x\in\mathbb{R}$, it follows that $S_c(\mu_t)-\widetilde{S}_c(\mu_t, t) = \alpha c (1+\tanh(t)) \geq \alpha c$, $t\geq 0$, a.s. Taking advantage of this together with Proposition 3.1.3, we can conclude that the following holds for the relative return of $\pi$ versus $\zeta$
\begin{align}
\label{master entropy comparison}
\log(Z_T^{\pi} / Z_T^{\zeta}) &= \log\left( \frac{\widetilde{S}_c(\mu_T, T) S_c(\mu_0)}{\widetilde{S}_c(\mu_0, 0) S_c(\mu_T)}  \right) + \frac{1}{2}\sum_{i=1}^n\int_0^T \frac{\alpha c (1+\tanh(t))}{\widetilde{S}_c(\mu_t, t) S_c(\mu_t)}\tau_t^{ii}\mu_t^i dt \nonumber \\ &+ \alpha c \int_0^T \frac{1-\tanh^2(-t)}{\widetilde{S}_c(\mu_t,t)}dt, \,\, T\geq 0, \,\, \text{a.s.},
\end{align}
where we have also used the drift process \eqref{entropyDrift} from Example 2.3.3. It also holds that
\begin{align}
\frac{\widetilde{S}_c(\mu_T, T) S_c(\mu_0)}{\widetilde{S}_c(\mu_0, 0) S_c(\mu_T)} &= \frac{S_c(\mu_T)+\alpha c(\tanh(-T)-1)}{S_c(\mu_T)} \cdot \frac{S_c(\mu_0)}{S_c(\mu_0)-\alpha c} \nonumber \\ &>
\frac{S_c(\mu_T)-2\alpha c}{S_c(\mu_T)} = 1-\frac{2\alpha c}{S_c(\mu_T)} \geq 1-2\alpha, \,\, T\geq 0, \,\, \text{a.s.},
\end{align}
where we have used $S_c(\mu_T) \geq c$, $T\geq 0,$ $\text{a.s.},$ in order to arrive at the last inequality. Note also that
\begin{equation}
\int_0^T \frac{1-\tanh^2(-t)}{\widetilde{S}_c(\mu_t, t)} dt \geq \frac{\tanh(T)}{c(1-\alpha)+\log(n)}, \,\, T\geq 0, \,\, \text{a.s.},
\end{equation}
similar to \eqref{integral estimate}. Using these observations in \eqref{master entropy comparison}, together with the bounds on $S_c$ and $\widetilde{S}_c$ and the weak sufficient intrinsic volatility assumption yields the following
\begin{align}
\label{master entropy comparison 2}
\log(Z_T^{\pi} / Z_T^{\zeta}) &> \log\left(1-2\alpha  \right) + \frac{\alpha c\Upsilon(T)}{(c(1-\alpha)+\log(n))(c+\log(n))} \nonumber \\ &+ \frac{\alpha c}{c(1-\alpha)+\log(n)}\tanh(T), \,\, T\geq 0, \,\, \text{a.s.}
\end{align}
By defining 
\begin{equation}
\Upsilon_c(T) = \Upsilon(T) + (c+\log(n))\tanh(T), \,\, T\geq 0,
\end{equation}
for all $c>0$, we deduce that on a time horizon $[0, T]$, where
\begin{equation}
T \geq  \widehat{T}_c(\alpha) := \Upsilon^{-1}_c \left( \frac{(c(1-\alpha)+\log(n))(c+\log(n))}{\alpha c}\log\left(\frac{1}{1-2\alpha}  \right)    \right),
\end{equation}
the portfolio $\pi$ outperforms the entropy weighted portfolio \eqref{entropyPortfolio} for all $c>0$, almost surely. Note that the inverse of $\Upsilon_c(T)$ with respect to $T$ exists as it is a strictly increasing continuous function. Furthermore, from
\begin{equation}
\widehat{T}_c = \lim_{\alpha \downarrow 0} \widehat{T}_c(\alpha) = \Upsilon_c^{-1}\left(\frac{2(c+\log(n))^2}{c} \right), \,\, c>0,
\end{equation}
we can infer that for a given $c>0$ and $T>\widehat{T}_c$, we can choose a small enough $\alpha>0$, such that the portfolio $\pi$ outperforms $\zeta$ on $[0, T]$. \QEDA\\
\\
{\bf Remarks.} Note that the previous example constructs a generalized generating portfolio which outperforms a strong relative arbitrage opportunity versus the market. Hence, we may conclude that considering stock characteristics processes in the context of the master equation can improve the performance of generating portfolios and that it can increase the outperformance with respect to the market.\\

Let us emphasize the contribution of the additional stock characteristics process in the previous two examples. As it is evident from the master equations \eqref{master reduction} and \eqref{master entropy comparison}, they generate a strictly positive, strictly increasing contribution to the relative return. Moreover, they improve the drift process, since the generating functions have a component which is strictly decreasing with time, and this leads to an amplification in \eqref{drift reduction} and \eqref{drift comparison}. However, this benefit comes with a trade-off, namely since $t \longmapsto \widetilde{S}(\cdot, t)$ is decreasing, it leads also to a smaller lower bound on $\log(\widetilde{S}(\mu_T, T)/\widetilde{S}(\mu_0, 0))$, which is the first term on the right hand side of the master equations. Let us remark that this is not an issue for the long-term performance of our portfolios, since the previous term is bounded, whereas the improved drift processes are unbounded and strictly increase with time. Thus, with a smart choice of the generating function and the stock characteristics process one can obtain portfolios which are more profitable than classical SPT strategies.
\subsection{Preference Based Investing in the Framework of SPT}
In this subsection we want to give two examples in order to illustrate how quantitative preferences can be included in making investment decisions  in our framework in order to outperform the market. 

In our first example we construct a portfolio based on market beta. Specifically, the trading strategy invests a larger amount of cash into stocks which have a smaller absolute correlation to the market portfolio. This is an inspiration from \cite{FP14}, where a ``betting against beta" factor is constructed in order to explain equity returns.\\
\\
{\bf 3.3.1 Example. (Portfolio selection based on market beta)} The beta of a stock is a key ingredient of the Capital Asset Pricing Model in order to explain equity returns. It measures the correlation between the returns of a stock and the returns of the market portfolio. In this paper, we shall define the {\it beta} process of a stock as
\begin{equation}
\beta_t^i := \left< X^i, Z^{\mu}\right>_t, \,\, t\geq 0,
\end{equation}
for $i=1,...,n$. Applying Ito's lemma on the process $\exp(\log(X^i))$ and using definition \eqref{model}, we obtain
\begin{equation}
dX_t^i = \left(\gamma_t^i +\frac{1}{2}\sum_{i=1}^n (\xi^{ij}_t)^2  \right)X_t^idt + X_t^i \sum_{j=1}^n \xi_t^{ij}dB_t^j, \,\, t\geq 0, \,\, \text{a.s.},
\end{equation}
for $i=1,...,n$. This observation, together with $Z^{\mu}=X^1+...+X^n$ gives
\begin{align}
d\beta_t^i = d\left< X^i, Z^{\mu} \right>_t &= X_t^i\sum_{j=1}^n \sum_{k=1}^n\sum_{\ell=1}^nX_t^k\xi^{ij}_t \xi^{k\ell}_t d\left< B^j, B^{\ell}\right>_t = X_t^i \sum_{k=1}^n \sigma_t^{ik} X_t^k dt, \,\, t\geq 0, \,\, \text{a.s.},
\end{align}
for $i=1,...,n$, where we have used the independence of Brownian motions and the definition of the covariance process \eqref{covariance}, in order to arrive at the last equality. 

We make the following two assumptions:
\begin{itemize}
\item[{\it i)}] The market $\mathfrak{M}$ is sufficiently volatile, i.e. there exists an $\varepsilon>0$, such that
\begin{equation}
\gamma_t^{\mu, *} \geq \varepsilon, \,\, t\geq 0, \,\, \text{a.s.}
\end{equation}
\item[{\it ii)}] The quantities 
\begin{equation}
\label{beta sign}
s^i_t := \text{sign}\left( X_t^i \sum_{k=1}^n \sigma_t^{ik} X_t^k  \right), \,\, t\geq 0, \,\, i=1,...,n,
\end{equation}
do not change over time, i.e. $s^i_t = s^i$, with $s^i \in \lbrace +1, -1 \rbrace$, for $t\geq 0$ and $i=1,...,n$, almost surely. In \eqref{beta sign}, sign$(x)=+1$, if $x\geq 0$, and $-1$ otherwise. Interpretation: once a stock is positively or negatively correlated to the market, it remains such for all times. This is motivated by empirical evidence which shows that stocks usually do not change the sign of ``their beta".
\end{itemize}
We shall consider the processes $\widetilde{\beta}^i := s^i \beta^i$, for $i=1,...,n$, as our stock characteristics process. We define the generating function $S_{A,c, p}$ by
\begin{equation}
S_{A,c,p}(x,y) := A + \sum_{i=1}^n (x^i)^p (c+\text{e}^{- y^i}), \,\, x=(x^1,...,x^n) \in \bigtriangleup_+^n, \,\, y=(y^1,...,y^n) \in [0, \infty)^n,
\end{equation}
for all $A\geq 0$, $c>0$ and $p\in(0,1)$. Obviously, $S_{A,c,p}\in\mathfrak{G}_{n}^{[0,\infty)^n}$. By taking advantage of Theorem 3.1.2, we conclude that the portfolio generated by $S_{A,c}$ reads
\begin{align}
\label{betaportfolio}
\pi_t^{i} &= \mu_t^i \left( \frac{ p({\mu_t^i})^{p-1}(c+\text{e}^{- \widetilde{\beta}_t^i})}{S_{A,c, p}(\mu_t, \widetilde{\beta}_t)}+ 1 - \frac{p}{S_{A, c,p}(\mu_t, \widetilde{\beta}_t)}\sum_{i=1}^n ({\mu_t^i})^p (c+\text{e}^{- \widetilde{\beta}_t^i})   \right) \nonumber \\ &= \frac{p({\mu_t^i})^p(c+\text{e}^{- \widetilde{\beta}_t^i}) + \mu_t^i ((1-p)S_{A,c,p}(\mu_t, \widetilde{\beta}_t)+ pA)}{S_{A,c,p}(\mu_t, \widetilde{\beta}_t)}, \,\, t\geq 0, \,\, i=1,...n.
\end{align}
It is evident that $\pi$ is a long-only portfolio for any $A\geq 0$, $c>0$ and $p\in(0,1)$. In addition, it allocates a larger amount of wealth into stocks which have a smaller absolute correlation to the market portfolio. We refer to it as the {\it beta weighted portfolio}. The corresponding drift process reads
\begin{equation}
d\Theta_t = \frac{p(1-p)}{2S_{A,c,p}(\mu_t, \widetilde{\beta}_t)}\sum_{i=1}^n (\mu_t^i)^p(c+\text{e}^{-\widetilde{\beta}_t^i})\tau_t^{ii}dt, \,\, t\geq 0, \,\, \text{a.s.}
\end{equation}
Let us also note that the generating function $S_{A,c,p}$ admits the following bounds
\begin{equation}
\label{s beta}
A+c \leq S_{A, c,p}(x,y) \leq A+(1+c)n^{1-p}, \,\, (x,y) \in \bigtriangleup_+^n \times [0,\infty)^n.
\end{equation}
The lower bound in \eqref{s beta} follows from $\sum_{i=1}^n ({x^i})^p \geq 1$, $x\in\bigtriangleup_+^n$, $p\in (0,1)$. In order to arrive at the upper bound of $S_{A,c,p}$, we have used the fact that $z\longmapsto z^p$, $z>0$, is concave for $p\in (0,1)$, and hence the inequality $z^p \leq z_0^p + pz_0^{p-1}(z-z_0)$ holds for all $z_0>0$. Applying this result with $z_0 = 1/n$ gives for all $x\in\bigtriangleup_+^n$
\begin{equation}
\sum_{i=1}^n (x^i)^p \leq \sum_{i=1}^n \left( \frac{1}{n^p} + \frac{p}{n^{1-p}}(x^i-1/n)  \right) = n^{1-p}.
\end{equation}
Thanks to Theorem 3.1.2, the relative return of the beta weighted portfolio with respect to the market satisfies
\begin{align}
\label{master_beta}
\log(Z_T^{\pi}/Z_T^{\mu}) &= \log(S_{A,c,p}(\mu_T, \widetilde{\beta}_T) / S_{A,c,p}(\mu_0, \widetilde{\beta}_0)) + \sum_{i=1}^n \int_0^T  \frac{p(1-p)}{2S_{A,c,p}(\mu_t, \widetilde{\beta}_t)}({\mu_t^i})^p(c+\text{e}^{-\widetilde{\beta}_t^i})\tau_t^{ii} dt \nonumber \\&+ \sum_{i=1}^n\int_0^T \frac{({\mu_t^i})^p \text{e}^{-\widetilde{\beta}_t^i}}{S_{A,c,p}(\mu_t, \widetilde{\beta}_t)}d\widetilde{\beta}_t^i, \,\, T\geq 0, \,\, \text{a.s.}
\end{align}
Note that the last sum in \eqref{master_beta} is positive and increasing almost surely, and hence improves the return of the beta weighted portfolio relative to the market. Using the upper and lower bound on $S_{A,c,p}$, the fact that $a^p> a$, for $a, p\in (0,1)$, as well as the positivity of the last term of \eqref{master_beta}, we arrive at the following estimate
\begin{align}
\log(Z_T^{\pi}/Z_T^{\mu}) &> \log\left(\frac{A+c}{A+(1+c){n}^{1-p}}\right) + \frac{p(1-p)c}{A+(1+c)n^{1-p}}\int_0^T \underbrace{\frac{1}{2}\sum_{i=1}^n\mu_t^i \tau_t^{ii}}_{=\gamma_t^{\mu, *}}dt \nonumber \\ &\geq \log\left(\frac{A+c}{A+(1+c){n}^{1-p}}\right) + \frac{p(1-p)c}{A+(1+c)n^{1-p}} \varepsilon T, \,\, T\geq 0,\,\, \text{a.s.},
\end{align}
where we have made usage assumption {\it ii)} in order to arrive at the last inequality.
Hence, on a time horizon $[0,T]$, where 
\begin{equation}
T \geq  \frac{A+(1+c)n^{1-p}}{p(1-p)c \varepsilon} \log\left(\frac{A+(1+c){n}^{1-p}}{A+c}\right),
\end{equation}
the beta weighted portfolio outperforms the market almost surely. \QEDA\\
\\
{\bf Remarks.} In the previous example we make the assumption that the market $\mathfrak{M}$ is sufficiently volatile and that the beta process of each company is either increasing or decreasing. Let us remark that such market models indeed exist. Consider the volatility stabilized model from \cite{FK05}, where the dynamics of stock prices are given by
\begin{equation}
\label{volastab}
d\log(X_t^{i}): = \frac{\alpha}{2\mu_t^i}dt + \frac{1}{\sqrt{\mu_t^i}}dB_t^i ,\,\,  t \geq 0 ,\,\, i=1,...,n,
\end{equation}
for an $\alpha\geq 0$. In terms of the notation used in \eqref{model}, we have that $\gamma^i = \alpha / (2\mu^i)$ and $\xi^{ij}= \delta^{ij}/\sqrt{\mu^i}$, for $i,j=1,...,n$. From this we can conclude that the covariance process $\sigma$ fulfils $\sigma_t^{ij} = \delta^{ij}/\mu_t^{i}$, for all $t\geq 0$ and $i,j=1,...,n$. A simple calculation shows then that the excess growth rate of the market is given by
\begin{equation}
\gamma_t^{\mu, *} = \frac{n-1}{2} ,\,\, t\geq 0  ,\,\, \text{a.s.}
\end{equation}
Hence, by Definition 2.3.5, the market model \eqref{volastab} is sufficiently volatile. Let us also remark that
\begin{equation}
\label{betaVSM}
d\beta^i_t = X_t^i \sum_{k=1}^n \sigma_t^{ik} X_t^k = \frac{(X_t^i)^2}{\mu_t^i} dt, \,\, t\geq 0, \,\, i=1,...,n, \,\, \text{a.s.}
\end{equation}
From \eqref{betaVSM} we are able to conclude that $s_t^i=+1$, $t\geq 0$, for all $i=1,...,n$, almost surely. Thus, volatility stabilized models satisfy both assumptions made in the previous example.\\

In the next example we construct a portfolio based on the performance metric {\it return on assets} (ROA). This metric tells how good a company is investing its assets in order to generate profit. Informally, we would expect companies which possess a smaller ROA to be less profitable than firms with a larger ROA, which is an idea from {\it quality investing}. Inspired by this, we construct a functionally generated portfolio which invests a larger amount of cash into stocks with a smaller ROA, and show that it underperforms the market portfolio with probability one, after a sufficiently long time horizon. Then using this observation, we construct an long-only portfolio which outperforms the market.\\
\\
{\bf 3.3.2 Example. (Portfolio selection based on ROA)} Let us introduce the $\mathbb{F}$-adapted, continuous-path semimartingales $R^1,...,R^n$, which represent the ROA processes of the companies considered in $\mathfrak{M}$. Specifically, $R_t^i$ is the ROA of company $i\in \lbrace 1,...,n \rbrace$ at time $t\geq 0$. Instead of giving an explicit model for $R^1,...,R^n$, we stay descriptive in our approach. We assume the following:
\begin{itemize}
\item[{\it i)}] The market weights are non-constant for all times and satisfy the {\it non-failure} condition, i.e. there exists a $\delta\in (0, 1/n) $ such that
\begin{equation}
\mu_t^i \geq \delta, \,\,t\geq 0, \,\,  i=1,...,n, \,\, \text{a.s.}
\end{equation}
Interpretation: There exists movement in the financial market and no company in $\mathfrak{M}$ can go bankrupt.
\item[{\it ii)}] There exists an $\varsigma >0$ such that $\mathsf{P}(0<R_t^{i}<\varsigma)=1$, for $i=1,...,n$ and $t\geq 0$. Interpretation: the ROA processes are bounded from above and below by a deterministic constant and companies cannot have a negative or arbitrary high ratio of net income to total assets.
\item[{\it iii)}] The quadratic covariation process between the market weights and the ROA processes vanishes for all companies, i.e. $\mathsf{P}(\left<\mu^{i}, R^i \right>_t \equiv 0)=1$, for $i=1,...,n$ and $t\geq 0$. Interpretation: the market capitalization of a company has no direct influence on its net income or assets (note that this might not be true for financial firms).
\item[{\it iv)}] There exist $\mathbb{F}$-adapted, continuous-path processes $\widetilde{R}^1,...,\widetilde{R}^n$ such that
\begin{equation}
d\left< R^i\right>_t = \widetilde{R}_t^i dt, \,\, t\geq 0, \,\, i=1,...,n, \,\, \text{a.s.}
\end{equation}
\begin{equation}
\sum_{i=1}^n \widetilde{R}_t^i \geq \eta >0, \,\, t>0, \,\, \text{a.s.}
\end{equation}
Interpretation: the ROA processes are ``sufficiently volatile".
\item[{\it v)}] There exist constants $A, \varepsilon\geq 0$, with $\varepsilon<\delta \text{e}^{-\varsigma}\eta /2$, such that for all $T<\infty$, the following holds for the market weights and the ROA processes
\begin{equation}
\label{assumption4}
\mathsf{P}\left(\sum_{i=1}^n\int_0^T  \mu_t^{i} \text{e}^{-R_t^{i}}dR_t^{i} <A+\varepsilon T \right)=1.
\end{equation}
Interpretation: the market $\mathfrak{M}$ is ``ROA diverse", in the sense that ROA processes of different companies are not comonotonic, but rather some move upwards and other downwards at same times.
\end{itemize}
We consider the ROA processes of the companies in $\mathfrak{M}$ as our stock characteristics process. Let us define the generating function $S$ by
\begin{equation}
S(x, y):= \exp\left(\sum_{i=1}^nx^{i}\text{e}^{-y^{i}} \right), \,\, x=(x^1,...,x^n) \in \bigtriangleup_+^n, \,\, y=(y^1,...,y^n) \in (0, \varsigma)^n.
\end{equation}
It is easy to verify that $S\in \mathfrak{G}_n^{(0, \varsigma)^n}$ and that it generates the portfolio
\begin{equation}
\label{roa portfolio}
\pi_t^{i} = \mu_t^{i}\left( \text{e}^{-R_t^{i}} + 1 - \sum_{j=1}^n \mu_t^j \text{e}^{-R_t^{j}} \right),\,\, t \geq 0,\,\, i =1,...,n.
\end{equation}
The trading strategy \eqref{roa portfolio} invests a larger amount of wealth in stocks with a smaller ROA, and a smaller amount of wealth into stocks with a larger ROA. Hence, we would expect this portfolio to underperform the market on the long run. Note that this portfolio is long-only. Indeed, since 
\begin{equation}
\label{4237}
\exp(-\varsigma) \leq \exp(-R_t^{i}) \leq 1, \,\, i=1,...,n, \,\, t\geq 0, \,\, \text{a.s.},
\end{equation}
we have for $i=1,...,n$
\begin{equation}
\pi_t^{i} \geq \mu_t^{i}\left( \text{e}^{-R_t^{i}} + 1 - \sum_{j=1}^n\mu_t^{j}\right) = \mu_t^{i}\text{e}^{-R_t^{i}} >0 ,\,\, t \geq 0,\,\, \text{a.s.}
\end{equation}
Next, let us look at the drift process generated by $S$. Recall that it is determined by
\begin{align*}
d \Theta_t &= - \frac{1}{2S(\mu_t, R_t)}\sum_{i, j =1}^n  \partial^{ij}S(\mu_t, R_t)\mu_t^{i}\mu_t^j \tau_t^{ij}dt  - \frac{1}{2}\sum_{i, j=1}^{k} \partial^{n+i,n+j} \log(S(\mu_t, R_t)) d\left< R^{i}, R^j\right>_t\nonumber  \\&- \sum_{i=1}^n\sum_{j=1}^{k} \partial^{i,n+j} \log(S(\mu_t, R_t)) d\left< \mu^{i}, R^j\right>_t ,\,\, t \geq 0,\,\, \text{a.s.}
\end{align*}
Thanks to the expression for the generating function and assumption $iii)$ on the ROA process, the expression for the drift process reads
\begin{equation}
\Theta_T = -\frac{1}{2}\sum_{i,j=1}^n\int_0^T \text{e}^{-R_t^{i}}\text{e}^{-R_t^{j}}\mu_t^{i} \mu_t^{j} \tau_t^{ij}   dt - \frac{1}{2}\int_0^T \sum_{i=1}^n \mu_t^{i}\text{e}^{-R_t^{i}} d\left< R^{i}\right>_t, \,\, t\geq 0, \,\, \text{a.s.}
\end{equation}
Defining the $\mathbb{R}^n$-valued process $z = (z^1_t,...,z^n_t)_{t \geq 0}$ componentwise as $z_t^{i}= \text{e}^{-R_t^{i}} \mu_t^{i}$, $t\geq 0$, $i=1,...,n$, yields then for all $T \geq 0$, almost surely
\begin{equation}
\label{drift estimate roa}
\Theta_T = -\frac{1}{2}\int_0^T z_t \cdot \tau_t z_t dt  - \frac{1}{2}\int_0^T \sum_{i=1}^n \mu_t^{i}\text{e}^{-R_t^{i}} d\left< R^{i}\right>_t \leq -\frac{1}{2}\int_0^T \sum_{i=1}^n \mu_t^{i}\text{e}^{-R_t^{i}} d\left< R^{i}\right>_t,
\end{equation}
where we have used Proposition 2.1.6, namely that the matrix-valued relative covariance process $\tau_t$ is positive semidefinite for all $t \geq 0$. Furthermore, using assumption $i)$ and the estimate \eqref{4237} in the first inequality, and assumption $iv)$ in the second inequality we get the following upper bound on the drift process generated by $S$
\begin{align}
\label{upper bound drift roa}
&-\frac{1}{2}\sum_{i=1}^n\int_0^T  \mu_t^{i}\text{e}^{-R_t^{i}} d\left< R^{i}\right>_t \leq -\frac{\delta \text{e}^{-\varsigma}}{2}\int_0^T\sum_{i=1}^n  d\left<R^i\right>_t \nonumber \\ &\leq -\frac{\delta \text{e}^{-\varsigma}}{2} \int_0^T \eta dt=-\frac{\delta \text{e}^{-\varsigma}}{2} \eta  T,\,\, T \geq 0,\,\, \text{a.s.}
\end{align}
Now, we want to show that the portfolio \eqref{roa portfolio} underperforms the market after a sufficiently long time horizon. For this purpose we look at the relative return of $\pi$ with respect to $\mu$, which according to Theorem 3.1.2 reads
\begin{align}
\label{master roa}
\log(Z_T^{\pi}/Z_T^{\mu}) &= \log(S(\mu_T, R_T)/S(\mu_0, R_0)) - \sum_{i=1}^n\int_0^T  \partial^{n+i}\log(S(\mu_t, P_t)) dR_t + \Theta_T
\nonumber  \\& = \sum_{i=1}^n (\mu_T^{i} \text{e}^{-R_T^{i}}-\mu_0^{i} \text{e}^{-R_0^{i}} ) +  \sum_{i=1}^n \int_0^T \mu_t^{i}e^{-R_t^{i}}dR_t^{i} + \Theta_T, \,\, T\geq 0, \,\, \text{a.s.}
\end{align}
Note that \eqref{4237} and the fact that $\mathsf{P}(\mu_t\in\bigtriangleup_+^n, t\geq 0)=1$ can be used to conclude that an upper bound for the first sum in \eqref{master roa} is given by $1-\text{e}^{-\varsigma}$. With the help of this remark, assumption $v)$ and \eqref{drift estimate roa} and \eqref{upper bound drift roa} we find
\begin{equation}
\label{rr roa estimate}
\log(Z_T^{\pi}/Z_T^{\mu}) \leq 1-\text{e}^{-\varsigma} + A +\varepsilon T - \frac{\delta \text{e}^{-\varsigma}}{2}\eta T,\,\, T \geq 0,\,\, \text{a.s.}
\end{equation}
Hence, if
\begin{equation}
\label{roa time}
T > T^*:=\frac{2(1+A-\text{e}^{-\varsigma})}{ \delta\eta\text{e}^{-\varsigma}-2\varepsilon},
\end{equation}
the market overperforms the ROA weighted portolio over $[0,T]$ almost surely. \QEDA \\

In the next instance, we make usage of the result of Example 3.3.2 in order to construct a long-only portfolio based on ROA, which outperforms the market almost surely.\\
\\
{\bf 3.3.3 Example. (A quality portfolio)} Let the financial market $\mathfrak{M}$ and the stock characteristics process be as in Example 3.3.2. Moreover, let $\pi$ denote the portfolio \eqref{roa portfolio}. We consider a portfolio $\eta$ which initially invests $1+a$ dollars in $\mu$ and sells $a$ dollars of portfolio $\pi$. Here $a$ denotes the quantity
\begin{equation}
a:= \frac{\text{e}^{-b}}{2-\text{e}^{-b}-\text{e}^{-\varsigma}},
\end{equation}
where we have defined $b := 1+A>0$, and $A$ is the constant from \eqref{assumption4}, whereas $\varsigma>0$ is the upper bound on the ROA processes. Note that $a>0$. The value of the portfolio $\eta$ is given by
\begin{equation}
Z_t^{\eta} = (1+a)Z_t^{\mu}-aZ_t^{\pi}, \,\, t\geq 0,
\end{equation}
with $Z_0^{\eta}=1$. First, we show that $Z_t^{\eta}>0,$ for $t\geq 0$, a.s. Indeed, from \eqref{rr roa estimate} it follows that
\begin{equation}
\label{inequality example 333}
Z_t^{\mu}\geq Z_t^{\pi} \exp(-b), \,\, t\geq 0, \,\, \text{a.s.}
\end{equation} 
Using this inequality, we get
\begin{align}
Z_t^{\eta} &\geq Z_t^{\pi} (\text{e}^{-b}(1+a)-a) = Z_t^{\pi}\frac{\text{e}^{-b}-\text{e}^{-\varsigma-b}}{2-\text{e}^{-b}-\text{e}^{\varsigma}} >0, \,\, t\geq 0, \,\, \text{a.s.}
\end{align}
The portfolio weights of $\eta$ read
\begin{equation}
\label{eta portfolio}
\eta_t^i = \frac{1}{Z_t^{\eta}}\left((1+a)Z_t^{\mu}\mu_t^i-aZ_t^{\pi}\pi_t^i \right), \,\, t\geq 0, \,\, i=1,...,n.
\end{equation}
We shall refer to the trading strategy \eqref{eta portfolio} as the {\it ROA weighted portfolio}. In the sequel we proceed to show that $\eta$ is a long-only portfolio which outperforms the market over a sufficiently long time horizon. Note that for arbitrary $t\geq 0$ it holds that
\begin{equation}
\sum_{i=1}^n\eta_t^i = \frac{(1+a)Z_t^{\mu}}{Z_t^{\eta}}\sum_{i=1}^n\mu_t^i - \frac{aZ_t^{\pi}}{Z_t^{\eta}}\sum_{i=1}^n \pi_t^i = \frac{(1+a)Z_t^{\mu}-aZ_t^{\pi}}{Z_t^{\eta}} = 1.
\end{equation}
Thus, $\eta$ defines a portfolio. In order to show that $\eta$ is long-only, it is enough to prove that
\begin{equation}
\label{kappa inequality}
\kappa_t^i:=(1+a)Z_t^{\mu}\mu_t^i-aZ_t^{\pi}\pi_t^i =\frac{(2-\text{e}^{-\varsigma})Z_t^{\mu}\mu_t^i - \text{e}^{-b}Z_t^{\pi}\pi_t^i}{2-\text{e}^{-b}-\text{e}^{-\varsigma}}\geq 0,
\end{equation} 
for all $t\geq 0$ and $i=1,...,n$, almost surely. By taking advantage of \eqref{inequality example 333} along with the definition $\kappa_t^i$, we get
\begin{equation}
\label{kappa estimate}
\kappa_t^i \geq \frac{Z_t^{\pi}\mu_t^i\text{e}^{-b}}{2-\text{e}^{-b}-\text{e}^{-\varsigma}}\left( 2-\text{e}^{-\varsigma}- \frac{\pi_t^i}{\mu_t^i} \right), \,\, t\geq 0, \,\, i=1,...,n, \,\, \text{a.s.}
\end{equation}
Moreover, the definition of the portfolio $\pi$ and \eqref{4237} imply
\begin{equation}
\pi_t^i \leq \mu_t^i(2-\text{e}^{-\varsigma}), \,\, t\geq 0, \,\, i=1,...,n, \,\, \text{a.s.}
\end{equation}
Using this finding in \eqref{kappa estimate} finally yields that $\kappa_t^i \geq 0$ for $t\geq 0$ and $i=1,...,n$, almost surely. Hence, $\eta$ is a long-only portfolio.

Next, we show that $\eta$ represents a strong relative arbitrage opportunity versus the market over $[0,T]$, for all $T> T^*$, where $T^*$ is given by \eqref{roa time}. Indeed, in Example 3.3.2 we have shown
\begin{equation}
Z_T^{\mu}>Z_T^{\pi}, \,\, T> T^*, \,\, \text{a.s.}
\end{equation}
Thus, we deduce that
\begin{equation}
Z_T^{\eta} = (1+a)Z_T^{\mu}-aZ_T^{\pi}>(1+a)Z_T^{\mu}-aZ_T^{\mu}=Z_T^{\mu},
\end{equation}
for all $T>T^*$, almost surely. \QEDA\\
\\
{\bf Remarks.} Note the significance of Example 3.3.3 as we end up with a strong relative arbitrage opportunity versus the market, which is generated by the volatility of the additional stock characteristics process. To the best of our knowledge it is the first such example across SPT literature, and it illustrates again the strength of the generalized setting we introduced in this section.
\section{Empirical Results}
In this section we report an empirical analysis of the trading strategies proposed in Section 3.3 and compare their performance to the market portfolio, the entropy weighted portfolio and the equally weighted portfolio. First, we describe our dataset along with the methodology of implementing the mentioned portfolios. Afterwards we present and visualize their performance statistics. Finally, we end this section with a regression analysis in order to understand the origins of the returns of our trading strategies.
\subsection{Datasets and Implementation}
The universe of stocks we consider consists out of $n=40$ firms. In particular we have considered the companies {\tt AAPL, AIG, AMGN, AXP, BA, BAC, C, CAT, CRM, CSCO, CVX, DD, DIS, GE, GS, HD, HON, HPQ, IBM, INTC, JNJ, JPM, KO, MCD, MDLZ, MMM, MO, MRK, MSFT, NKE, PFE, PG, RTX, T, TRV, UNH, VZ, WBA, WMT} and {\tt XOM}. Our testing period ranges from the 3rd of January 2006 till the 31st of December, 2020. We were given the access to the daily prices, historical shares outstanding and ROA data for the above stocks. Note that the latter two quantities are not reported on a daily basis, hence at each day in the time range we simply use the last reported value. Accounting data is usually published with a lag of several months, which is the reason why the dates in our shares outstanding and ROA dataset do not correspond to the dates the data was actually published. Let us also remark that we did not have access to historical data of delisted companies. The previous two remarks lead us to the conclusion that {\it survivorship bias} and {\it look-ahead bias} are present in our analysis. Due to these unavoidable biases, we can deduce that the quality of our datasets is rather poor. Finally, aiming to understand the sources driving the returns of our portfolios, we have used the daily historical returns of the three Fama French risk factors. These are available together with the historical risk-free rate on the website of Kenneth French\footnote{\url{https://mba.tuck.dartmouth.edu/pages/faculty/ken.french/data_library.html}}.

We rebalance all portfolios on a daily basis. The first portfolio for each strategy is formed on January 3rd, 2006, and the last portfolio is implemented on December 30th, 2020. This corresponds to a discrete time horizon $\lbrace 0,1,...,T\rbrace$ with $T=3775$. Each trade causes proportional transaction costs and an additional ``overnight fee" for portfolios which short stocks. We describe in detail the used model in the next subsection. All our calculations and analysis have been performed in the programming language $\mathsf{Python}$.

Let us note that the entropy weighted portfolio from Example 2.3.3 was computed with $c=10^{-1}$. In order to implement the beta weighted portfolio from Example 3.3.1, we have set $A=10^{-4}$, $c=10^{-4}$ and $p=0.7$. All these choices are motivated by the master equations of the corresponding portfolios, aiming to amplify the drift processes. Moreover, we have estimated the ``betas" as
\begin{equation}
\beta_t^i = \sum_{s=1}^t (X_s^i-X_{s-1}^i)(Z_s^{\mu}-Z_{s-1}^{\mu}), \,\, t=1,...,T,
\end{equation}
with $\beta_0^i=0$, for all $i=1,...,n$. We have also assumed $s^i=+1$, for $i=1,...,n$, regarding the beta weighted portfolio. This is indeed observed empirically in our dataset for a majority of time points. In order to implement the ROA weighted portfolio from Example 3.3.3 the value of $a$ was set to $2.5$. Note that the choice of $a$ is rather greedy and indeed results with the ROA weighted portfolio occasionally selling stocks. Let us remark that arbitrarily high values of $a$ would not result in a desirable portfolio since its value process would deplete quickly to 0 due to our transaction costs model which accounts for ``overnight fees" when a portfolio is short in a stock. Let us also comment that the ROA of a stock is a number relatively close to 0, and hence the portfolio \eqref{roa portfolio} will approximately be equal to the market portfolio. In order to prevent this, we scaled all ROA values by multiplying them with a factor of 10. We also want to stress that at no point, any optimization of the above parameters was performed. This is in order to reduce the likelihood of a {\it backtest overfit}. However, a rigorous search of parameters may be done on a dataset which possesses a larger time range. One could split the dataset into two subsamples, where one is used for choosing the optimal parameters (e.g. by a simple grid search algorithm) and the other for performing an out-of-sample test.
\subsection{Portfolio Performance}
In Figure 1 we present the portfolio value processes of the market portfolio (Market), the entropy weighted portfolio from Example 2.3.3 (Entropy), the equally weighted portfolio from Example 3.1.4 (EWP), the beta weighted portfolio from Example 3.3.1 (Beta) and the ROA weighted portfolio from Example 3.3.3 (ROA). In Table 1, we listen the performance statistics of these portfolios and their corresponding value processes. In particular, for each portfolio we report the annualized return, the annualized Sharpe ratio, the annualized information ratio, terminal value of the portfolio, annualized turnover, as well as the annualized ``alpha" and ``beta". In the sequel, we summarize how we have computed these performance measures and characteristics.\\

Let us fix some $\varepsilon_1, \varepsilon_2 \geq 0$. The {\it discrete portfolio value} process $Z^{\pi}=(Z_t^{\pi})_{t=0,...,T}$ is determined by
\begin{equation}
\label{wealth process}
Z_t^{\pi}:= Z_0 \prod_{s=1}^t (1+R_s^{\pi}), \,\, t=1,...,T,
\end{equation}
and $Z_0^{\pi}:=Z_0$, for any portfolio $\pi = (\pi_t)_{t=0,...,T-1}$. The quantity $Z_0$ represents the initial fortune of an investor, which we have normalized to 1 for all portfolios and $R^{\pi} = (R_t^{\pi})_{t=1,...,T}$ is the {\it portfolio return} process given as
\begin{equation}
\label{portfolio return}
R_t^{\pi} := {\sum_{i=1}^n \pi^i_{t-1}\left(\frac{X^i_t}{X^i_{t-1}}-1\right)} - \varepsilon_1 \sum_{i=1}^n|\pi_{t-1}^i-\hat{\pi}_{t-1}^i| - \varepsilon_2 \sum_{i=1}^n \pi_{t-1}^{\downarrow, i}, \,\, t=1,...,T.
\end{equation}
In \eqref{portfolio return}, the second sum captures the effect of proportional transaction costs and $\hat{\pi} = (\hat{\pi}_t^1,...,\hat{\pi}_t^n)_{t=0,...,T-1}$ are the {\it readjusted portfolio weights} given by
\begin{equation}
\hat{\pi}_t^i := \pi_{t-1}^i \frac{X_t^i}{X_{t-1}^i}\frac{Z_{t-1}^{\pi}}{Z_t^{\pi}}, \,\, t=1,..., T-1,
\end{equation}
and $\hat{\pi}_0^i =0$ for $i=1,...,n$. The third sum represents the additional costs caused by portfolios which short stocks. Specifically, $\pi^{\downarrow}=(\pi_t^{\downarrow,1},...,\pi_t^{\downarrow,n})_{t=0,...,T-1}$ is the ``{\it short leg}" of $\pi$ determined by
\begin{equation}
\pi_t^{\downarrow, i} := \max (0, -\pi_t^i), \,\, t=0,...,T-1, \,\, i=1,...,n.
\end{equation}
We have set $\varepsilon_1 =0.3 \%$ and $\varepsilon_2=0.5\%$ in our numerical experiments. Furthermore, the {\it annualized return} of a portfolio is calculated according to
\begin{equation}
\overline{R^{\pi}} := \left(\frac{Z_T^{\pi}}{Z_0}\right)^{252/T}-1.
\end{equation}
For realizations $X_1,...,X_m$ of a random variable $X$, we denote by $\widehat{\mathsf{E}}[X]$ and $\hat{\sigma}(X)$ its empirical mean and standard deviation, respectively. In particular
\begin{equation}
\widehat{\mathsf{E}}[X] := \frac{1}{m}\sum_{i=1}^m X_m,
\end{equation}
\begin{equation}
\widehat{\sigma}(X) := \frac{1}{m-1} \sum_{i=1}^m (X_i-\widehat{\mathsf{E}}[X]).
\end{equation}
We define the {\it annualized Sharpe ratio} of a portfolio $\pi$ as
\begin{equation}
\mathsf{SR}(\pi) := \sqrt{252} \frac{\widehat{\mathsf{E}}[R^{\pi}]}{\widehat{\sigma}(R^{\pi})}.
\end{equation}
We also define the {\it annualized information ratio} of a portfolio $\pi$ with respect to the market portfolio as
\begin{equation}
\mathsf{IR}_{\mu}(\pi) := \sqrt{252} \frac{\widehat{\mathsf{E}}[R^{\pi}-R^{\mu}]}{\widehat{\sigma}(R^{\pi}-R^{\mu})}.
\end{equation}
We use the following equation to determine the {\it annualized turnover} of $\pi$
\begin{equation}
\overline{\Delta^{\pi}} := \frac{252}{T}\sum_{t=0}^{T-1} \sum_{i=1}^n |\pi_{t}^i-\hat{\pi}_{t}^i|.
\end{equation}
Finally, we calculate the ``{\it alpha}" $\alpha^{\pi}$  and ``{\it beta}" $\beta^{\pi}$ of a portfolio $\pi$ as the regression coefficients in the model
\begin{equation}
\label{CAPM}
R_t^{\pi} - R_t^f := \alpha^{\pi} + \beta^{\pi} (R_t^{\mu}-R_t^f) + \zeta_t^{\pi}, \,\, t=1,..., T,
\end{equation}
where $R_t^f$ is the risk-free rate and $\zeta_1^{\pi},...,\zeta_T^{\pi}$ are i.i.d. with mean 0. The {\it annualized alpha} $\overline{\alpha^{\pi}}$ is then determined by
\begin{equation}
\label{annualized alpha}
\overline{\alpha^{\pi}} := 252\cdot \alpha^{\pi}.
\end{equation}
Note that the $\overline{\alpha^{\pi}}$ reported in Table 1 are expressed in percentages. In addition, we also listen the {\it coefficient of determination} $\mathsf{R}^2$ associated to the regression \eqref{CAPM}, for each portfolio. For the ROA weighted portfolio we have also computed the {\it mean amount shorted} $\overline{\pi^{\downarrow}}$ with help of the expression
\begin{equation}
\overline{\pi^{\downarrow}} = \frac{1}{T}\sum_{t=0}^{T-1} \sum_{i=1}^n \pi_t^{\downarrow, i},
\end{equation}
which resulted in $\overline{\pi^{\downarrow}} = 0.005$.
\begin{figure}[H]
\centering\includegraphics[width=15cm]{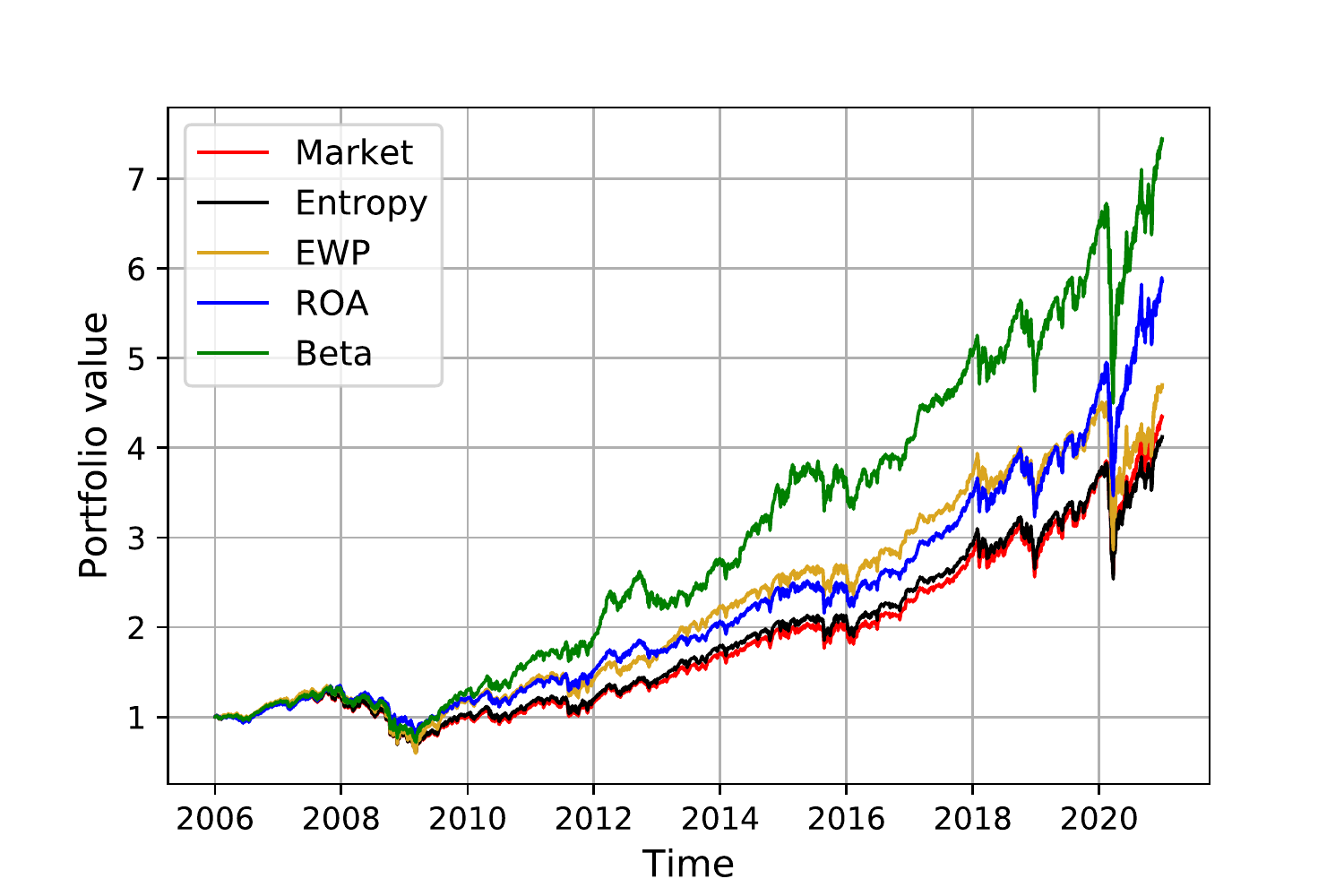}
\caption{The transaction cost adjusted portfolio value processes of the market portfolio (Market), the entropy weighted portfolio (Entropy), the equally weighted portfolio (EWP), the beta weighted portfolio from Example 3.3.1 (Beta) and the ROA weighted portfolio from Example 3.3.3 (ROA). The testing period starts on the 3rd of January, 2006 and ends with the 31st of December, 2020.}
\label{fig1}
\end{figure}
\begin{table}[H]
\centering
\begin{tabular}{ c c c c c c c c c}
\hline
 \hline
  Portfolio &$\,\, \overline{R^{\pi}} \,\, $ & $\,\, \mathsf{SR}(\pi)\,\,$ &$\,\, \mathsf{IR}_{\mu}(\pi)\,\, $ & $\,\,\,\, Z_T^{\pi}\,\,\,\, $ & $\,\, \overline{\Delta^{\pi}}\,\,$ & $\,\, \overline{\alpha^{\pi}}\,\, $ & $\,\, \beta^{\pi}\,\, $& $\,\, \mathsf{R}^2\,\, $ \\
 \hline
 \hline
 Market & 10.30\% & 0.596  & - &4.345 & 0.126 &0 &1 & 1\\
\hline
Entropy &9.92\%  & 0.577 &-0.218 &4.124 & 0.706 &-0.371 &1.000&0.995\\
\hline
EWP  &10.88\%  &0.600  & 0.167 &4.703 &2.338 & 0.253 &1.031 &0.958\\
\hline
ROA  & 12.52\% & 0.724  &0.420 &5.856 & 0.414 & 2.686 & 0.927&0.955\\
\hline
Beta  & 14.33\% & 0.787 & 0.612 & 7.442 & 0.813 & 4.124 & 0.941 & 0.915 \\
\hline
\hline

 \end{tabular}
 \caption{Performance statistics and portfolio characteristics of the tested strategies from Figure 1. For a portfolio $\pi$, the quantity $\overline{R^{\pi}}$ denotes the annualized return, $\mathsf{SR}(\pi)$ is the annualized Sharpe ratio, $\mathsf{IR}_{\mu}(\pi)$ is the annualized information ratio with respect to the market portfolio, $Z_T^{\pi}$ represents the portfolio value on terminal time (31st of December, 2020), $\overline{\Delta^{\pi}}$ is the annualized turnover, $\overline{\alpha^{\pi}}$ and $\beta^{\pi}$ are the annualized alpha and market beta respectively, and $\mathsf{R}^2$ is the coefficient of determination associated to the regression model \eqref{CAPM}.}
\end{table}

From Figure 1 and Table 1 we can observe that the ROA weighted and beta weighted portfolios indeed have the potential to be desirable investment strategies. Not only do they outperform the considered SPT alternatives pathwise, they also admit a higher Sharpe ratio, which means that they deliver a higher return for the same level of risk undertaken. Moreover, the generalized generating portfolios record high information ratios, larger alphas and smaller betas compared to the entropy weighted portfolio and the EWP. This allows us to conclude that they are able to outperform the market significantly on a consistent basis, which is also confirmed visually in Figure 1.
\subsection{Sources of Return}
In order to understand the factors driving the returns of our portfolios, we have performed regressions on the three Fama French risk factors (see \cite{FF92} and \cite{FF93}). Specifically, we have estimated the regression coefficients $\alpha^{\pi}$, $\beta^{\pi}$, $\mathfrak{s}^{\pi}$ and $\mathfrak{h}^{\pi}$ in the model
\begin{equation}
\label{fama french}
R_t^{\pi}-R_{t}^f := \alpha^{\pi} + \beta^{\pi} (R_t^{\mu}-R_t^f) + \mathfrak{s}^{\pi} R_t^{\mathfrak{s}} +\mathfrak{h}^{\pi} R_t^{\mathfrak{h}}+ \zeta_t^{\pi}, \,\, t=1,...,T.
\end{equation}
In \eqref{fama french}, $R_t^{\mathfrak{s}}$ is the return of the ``Small Minus Big" (SML) size factor and $R_t^{\mathfrak{h}}$ is the return of the ``High Minus Low" (HML) value factor. In Table 2 we report the annualized alpha $\overline{\alpha^{\pi}}$, the ``factor loadings" $\beta^{\pi}$, $\mathfrak{s}^{\pi}$, $\mathfrak{h}^{\pi}$,  as well as the $\mathsf{R}^2$ for the regression model \eqref{fama french}, for all portfolios. 
\begin{table}[H]
\centering
\begin{tabular}{  c c c c c c  }
\hline
 \hline
  Portfolio & $\,\,\,\,\,\,\, \overline{\alpha^{\pi}} \,\,\,\,\,\,\,\ $  & $\,\,\,\,\,\,\, \beta^{\pi} \,\,\,\,\,\,\,\ $  & $\,\,\,\,\,\,\,\ \mathfrak{s}^{\pi}\,\,\,\,\,\,\,\ $ & $\,\,\,\,\,\,\,\ \mathfrak{h}^{\pi}\,\,\,\,\,\,\,\ $ & $\,\,\,\,\,\,\,\ \mathsf{R}^2\,\,\,\,\,\,\,\ $ \\
 \hline
 \hline
 Market& 0.422 & 0.963 & -0.256 & 0.006 &0.970 \\
\hline
Entropy & 0.408  & 0.954 & -0.230 & 0.074 &0.974   \\
\hline
EWP  & 1.669  & 0.969  & -0.170 &0.200 &0.967   \\
\hline
ROA  & 1.710 & 0.935  &-0.261 & -0.238 &0.936   \\
\hline
Beta  & 4.072 & 0.914 & -0.204 & -0.088 & 0.870  \\
\hline
\hline
 \end{tabular}
 \caption{The loadings on the three Fama French risk factors of the returns of our portfolios. Specifically, for portfolio $\pi$, $\overline{\alpha^{\pi}}$ is the annualized alpha, $\beta^{\pi}$ is the loading on the market risk factor, $\mathfrak{s}^{\pi}$ is the loading on SMB and $\mathfrak{h}^{\pi}$ is the exposure on HML. The quantity $\mathsf{R}^2$ is the coefficient of determination associated to the regression model \eqref{fama french}.} 
\end{table}
From Table 2 it is evident that all our portfolios load positively on the market risk factor and negatively on the SMB risk factor. This was indeed expected in beforehand as our investment universe consists mainly of large capitalization stocks which tend to have high weights in the overall market portfolio. Moreover, we observe that the returns of the market portfolio, entropy weighted portfolio and EWP load positively on the value factor, whereas the returns of the ROA weighted portfolio and the beta weighted portfolio have a negative exposure on HML. In addition, the ROA weighted portfolio and the beta weighted portfolio show the largest annualized alpha. However, the regressions related to the latter two portfolios also possess the lowest $\mathsf{R}^2$. 

Even though the model \eqref{fama french} manages to explain the returns of our portfolios accurately, as all $\mathsf{R}^2$ are higher than 0.87, it would be interesting to examine the factor exposures of our portfolios in other models as well. In particular, we have in mind the Carhart four factor model, which is an extension of \eqref{fama french} by an additional term which describes momentum (see \cite{C97}). 
\section{Discussion}
In this paper we have further generalized the master equation originally introduced in \cite{F02}. While there have been already attempts in doing so, by means of adding a process of finite variation as the argument of the generating function, our approach goes a step further as we allow generating functions to depend on continuous-path semimartingales, in addition to the market weights. To the best of our knowledge, this is the first paper which makes a step in this direction and in addition explicitly demonstrates the value of a generalized master equation. We have shown that it is possible to shorten time horizons beyond which relative arbitrage is possible, increase the outperformance of generating portfolios with respect to the market and include preference based investing in the framework of SPT. Semimartingales represent a more interesting class of stock characteristics than finite variation processes, as the former can enhance the drift process significantly. Furthermore, they allow for a greater modelling flexibility, as one is able to drop potentially unrealistic assumptions on the volatility of the market weights, and replace them by more realistic assumptions on the stock characteristics process, in order to generate relative arbitrages. However, this comes then with a cost of having a stochastic integral in the master equation.

Stochastic portfolio theory still offers a rich amount of research possibilities. The main open problem is the incorporation of transaction costs within the master equation. Transaction costs can affect the returns of a portfolio significantly and it would be of great value to incorporate them in the context of SPT. An empirical study of the impact of proportional transaction costs on some functionally generated portfolios is given in \cite{RX20}. From a practitioner's point of view, it would be interesting to rigorously examine the validity of SPT results in discrete time. This was pointed out in \cite{V15}. From the standpoint of our work it would be appealing to question the existence of short-term relative arbitrage opportunities, possibly generated by the additional stock characteristics process.  Moreover, it would be interesting to further examine stock characteristics processes and generating functions which amplify the drift process in the generalized master equation, and possibly look for portfolios which beat the market on average, instead of almost surely. To search for models which fulfil $\mathsf{E}[\log(Z_T^{\pi}Z_T^{\mu})] \geq 0$ (where $\mathsf{E}[\cdot]$ denotes the expected value under the probability measure $\mathsf{P}$) would allow flexibility in handling the stochastic integral from the master equation, while maintaining the benefits of the improved drift process. 

\numberwithin{equation}{section}
\section*{Appendix: Proof of Theorem 3.1.2}
The weights given by \eqref{portfolio weights} sum to 1. Furthermore, they are bounded and $\mathbb{F}$-progressively measurable. This implies that $\pi$ is a portfolio. The process $\Theta$ given by \eqref{drift process} is clearly of finite variation. \\

Let $P$ be a $K$-valued stock characteristics process and $S\in \mathfrak{G}_n^K$. We start by applying Ito's lemma on $\log(S(\mu, P))$
\begin{align}
\label{ito lemma on S}
d \log(S(\mu_t, P_t))&=\sum_{i=1}^n \partial^{i}\log(S(\mu_t, P_t)) d\mu_t^{i} + \sum_{i=1}^{k} \partial^{n+i}\log(S(\mu_t, P_t)) dP^{i}_t \nonumber  \\&+ \frac{1}{2}\sum_{i,j=1}^n \partial^{ij}\log(S(\mu_t, P_t))d\left <\mu^{i}, \mu^j\right>_t + \frac{1}{2}\sum_{i,j=1}^k \partial^{n+i,n+j}\log(S(\mu_t, P_t))d\left <P^{i}, P^j\right>_t
\nonumber \\&+  \sum_{i=1}^n \sum_{j=1}^k \partial^{i,n+j}\log(S(\mu_t, P_t)) d\left<\mu^{i}, P^{j}\right>_t,\,\, t \geq 0 ,\,\,  \text{a.s.},
\tag{A.1}
\end{align}
where we take advantage of Schwarz's theorem and the symmetry of $\left<\cdot, \cdot \right>$ to deduce that the last term in \eqref{ito lemma on S} lacks a factor $1/2$. Since $S$ generates a portfolio $\pi$ with stock characteristics $P$, we have from Definition 3.1.1
\begin{align}
\label{equality}
d\log(Z_t^{\pi}/Z_t^{\mu}) &= d\log(S(\mu_t, P_t)) -\sum_{i=1}^k \partial^{i}\log(S(\mu_t, P_t))dP_t^{i} + d\Theta_t,\,\, t \geq 0 ,\,\,  \text{a.s.}
\tag{A.2}
\end{align}
By inserting \eqref{ito lemma on S} into \eqref{equality} we get
\begin{align}
\label{local martingale}
d\log(Z_t^{\pi}/Z_t^{\mu})&=
\sum_{i=1}^n \partial^{i}\log(S(\mu_t, P_t)) d\mu_t^{i} + \frac{1}{2}\sum_{i,j=1}^n \partial^{ij}\log(S(\mu_t, P_t))d\left <\mu^{i}, \mu^j\right>_t \nonumber \\&+ \frac{1}{2}\sum_{i,j=1}^k \partial^{n+i,n+j}\log(S(\mu_t, P_t))d\left <P^{i}, P^j\right>_t
\nonumber \\&+ \sum_{i=1}^n \sum_{j=1}^k \partial^{i,n+j}\log(S(\mu_t, P_t)) d\left<\mu^{i}, P^{j}\right>_t  + d\Theta_t,\,\, t \geq 0 ,\,\,  \text{a.s.}
\tag{A.3}
\end{align}
Proposition 2.2.3 states that the relative return process of a portfolio $\pi$ versus the market satisfies for all $t\geq 0$, almost surely
\begin{equation}
\label{rel.return}
d\log(Z^{\pi}_t/Z_t^{\mu}) = \sum_{i=1}^n \pi_t^{i} d\log(\mu_t^{i}) + \frac{1}{2} \sum_{i=1}^n \pi_t^{i} \tau_t^{ii}dt - \frac{1}{2} \sum_{i,j=1}^n \pi_t^{i}\tau_t^{ij} \pi_t^{j}dt, \tag{A.4}
\end{equation}
where we have taken advantage of the numeraire invariance property (Proposition 2.1.7) to express the excess growth rate $\gamma^{\pi, *}$ in terms of the relative covariance process of the market $\tau$. By definition of the market weights, respectively the relative covariance process we have for $t\geq 0$, a.s.
\begin{equation}
\label{rel.covariance market weights}
d\left< \log(\mu^i), \log(\mu^{j}) \right>_t = d\left< \log(X^i/Z^{\mu}), \log(X^j/Z^{\mu}) \right>_t = \tau_t^{ij}dt. \tag{A.5}
\end{equation}
Furthermore, an application of Ito's lemma on $\mu_t^{i} = \exp(\log(\mu_t^{i}))$ yields
\begin{equation}
\label{dmu}
d\mu_t^{i} = \mu_t^{i} d\log(\mu_t^{i}) + \frac{1}{2} \mu_t^{i} \tau_t^{ii}dt,\,\, t \geq 0 ,\,\, i=1,...,n, \,\,  \text{a.s.}, \tag{A.6}
\end{equation}
where \eqref{rel.covariance market weights} was used. From \eqref{dmu} it also follows that 
\begin{equation}
\label{quadratic variation market weights}
d\left< \mu^{i}, \mu^{j}\right>_t = \mu_t^{i}\mu_t^{j}\tau_{t}^{ij}dt, \,\, i,j=1,...,n, \tag{A.7}
\end{equation}
for all $t\geq 0$, a.s. Expressing $d\log(\mu_t^{i})$ by means of \eqref{dmu} and inserting it into \eqref{rel.return} gives
\begin{equation}
\label{rel.return rewritten}
d\log(Z^{\pi}_t/Z_t^{\mu}) = \sum_{i=1}^n \frac{\pi_t^{i}}{\mu_t^{i}}d\mu_t^{i} - \frac{1}{2}\sum_{i,j=1}^{n}\pi_t^{i}\tau_t^{ij} \pi_t^{j} dt,\,\, t \geq 0 ,\,\,  \text{a.s.} \tag{A.8}
\end{equation}
In order for \eqref{equality} to hold, the local martingale parts of \eqref{local martingale} and \eqref{rel.return rewritten} have to be equal. This is indeed satisfied if
\begin{equation}
\label{portfolio theta}
\pi_t^{i}=\mu_t^{i}(\partial^{i} \log(S(\mu_t, P_t))+\vartheta_t), \,\, i = 1,...,n ,\,\, t\geq 0, \tag{A.9}
\end{equation}
for any $\mathbb{R}$-valued stochastic process $\vartheta = (\vartheta_t)_{t\geq 0}$. To see the above statement we remark that
\begin{align}
\sum_{i=1}^n\frac{\pi_t^{i}}{\mu_t^{i}}d\mu_t^{i} &= \sum_{i=1}^n(\partial^{i} \log(S(\mu_t, P_t))+\vartheta_t)d\mu_t^{i} =  \sum_{i=1}^n\partial^{i} \log(S(\mu_t, P_t))d\mu_t^{i} \tag{A.10}
\end{align}
holds for $t\geq 0$, a.s., since $\sum_{i=1}^n d\mu_t^{i}=0$. Hence, the local martingale parts of \eqref{local martingale} and \eqref{rel.return rewritten} are equal. The process $\vartheta$ is determined in such a way that \eqref{portfolio theta} defines a portfolio. It is easy to see that this is the case if $\vartheta_t$ is given by
\begin{equation}
\vartheta_t = 1 - \sum_{j=1}^n
\mu_t^j \partial^j \log(S(\mu_t, P_t)),\,\, t\geq 0 ,\,\, \text{a.s.} \tag{A.11}
\end{equation}
This proves \eqref{portfolio weights}. Now that we have equality of the local martingale parts of \eqref{local martingale} and \eqref{rel.return rewritten}, we also want the finite variation parts of the two equations to be equal. This requirement determines the drift process. In particular, by comparing \eqref{local martingale} and \eqref{rel.return rewritten}, we have for any $t\geq 0$, almost surely
\begin{align}
\label{drift process 2}
d\Theta_t &= - \frac{1}{2}\sum_{i,j=1}^{n}\pi_t^{i}\tau_t^{ij} \pi_t^{j} dt  - \frac{1}{2}\sum_{i,j=1}^n \partial^{ij}\log(S(\mu_t, P_t))\mu_t^i\mu_t^j \tau_t^{ij}dt \nonumber \\&- \frac{1}{2}\sum_{i,j=1}^k \partial^{n+i,n+j}\log(S(\mu_t, P_t))d\left <P^{i}, P^j\right>_t
\nonumber \\&-  \sum_{i=1}^n \sum_{j=1}^k \partial^{i,n+j}\log(S(\mu_t, P_t)) d\left<\mu^{i}, P^{j}\right>_t, \tag{A.12}
\end{align}
\\
where we have taken advantage of identity \eqref{quadratic variation market weights} in the first line. Using the obtained expression for $\pi_t^{i}$ it follows that
\begin{align}
\label{pitaupi}
\sum_{i,j=1}^{n}\pi_t^{i}\tau_t^{ij} \pi_t^{j} &= \sum_{i,j=1}^{n} \big( \mu_t^{i}(\partial^{i}\log(S(\mu_t, P_t)) + \vartheta_t) \big) \big( \mu_t^{j}(\partial^{j}\log(S(\mu_t, P_t)) + \vartheta_t)\big)\tau_t^{ij}  \nonumber \\&= \sum_{i,j=1}^n \partial^{i}\log(S(\mu_t, P_t)) \partial^{j}\log(S(\mu_t, P_t))\mu_t^{i}\mu_t^{j} \tau_t^{ij},\,\, t\geq 0,\,\, \text{a.s.}, \tag{A.13}
\end{align}
where we have used Proposition 2.1.6, which states that $\mu_t$ spans the kernel of $\tau_t$ for all $t \geq 0$ a.s. One can also verify that the following holds for all $x,y \in \bigtriangleup_+^n \times K$, $i,j=1,...,n$
\begin{equation}
\label{4118}
\partial^{ij}\log(S(x,y)) = \partial^{ij}S(x,y)/S(x,y) - \partial^{i} \log(S(x,y))\partial^{j }\log(S(x,y)). \tag{A.14}
\end{equation}
Using \eqref{pitaupi} in the first equality, along with \eqref{4118} in the second equality, we see that the first two terms on the right hand side of \eqref{drift process 2} satisfy
\begin{align}
& - \frac{1}{2}\sum_{i,j=1}^{n}\pi_t^{i}\tau_t^{ij} \pi_t^{j} dt  - \frac{1}{2}\sum_{i,j=1}^n \partial^{ij}\log(S(\mu_t, P_t))\mu_t^i \mu_t^j \tau_t^{ij} dt \nonumber \\ &= -\frac{1}{2}\sum_{i,j=1}^n \left(  \partial^{i}\log(S(\mu_t, P_t)) \partial^{j}\log(S(\mu_t, P_t)) +\partial^{ij} \log(S(\mu_t, P_t))  \right)\mu_t^{i}\mu_t^{j} \tau_t^{ij}dt\nonumber \\ &= -\frac{1}{2S(\mu_t, P_t)} \sum_{i,j=1}^n \partial^{ij}S(\mu_t, P_t)\mu_t^i \mu_t^j \tau_t^{ij}dt,\,\, t\geq 0,\,\, \text{a.s.}, \tag{A.15}
\end{align}
which finally gives the desired expression \eqref{drift process} for the drift process. \QEDA \\

\addcontentsline{toc}{section}{\protect\numberline{}References}%


\begin{thebibliography}{99}

\bibitem[C97]{C97} Mark Carhart, {\it On Persistence in Mutual Fund Performance}, The Journal of Finance, 52(1):57–82, 1997.

\bibitem[CE15]{CE15} Samuel Cohen and Robert Elliott, {\it Stochastic Calculus and Applications}, Birkh\"auser Verlag, 2015.

\bibitem[CEH16]{CEH16} Lawrence Cunningham, Torkell Eide and Patrick Hargreaves, {\it Quality Investing: Owning the best companies for the long term}, Harriman House, 2016.

\bibitem[EK19]{EK19} Ernst Eberlein and Jan Kallsen, {\it Mathematical Finance}, Springer, 2019.

\bibitem[F02]{F02} Robert Fernholz, {\it Stochastic Portfolio Theory}, Springer, 2002.

\bibitem[FF92]{FF92} Eugene Fama and Kenneth French, {\it The cross-section of expected stock returns}, The Journal of Finance, 47(2):427–465, 1992.

\bibitem[FF93]{FF93} Eugene Fama and Kenneth French, {\it Common risk factors in the returns on stocks and bonds}, Journal of Financial Economics, 33(1):3–56, 1993.

\bibitem[FKK05]{FKK05} Robert Fernholz, Ioannis Karatzas and Constantinos Kardaras, {\it Diversity and relative arbitrage in equity markets}, Finance Stoch., 9(1):1–27, 2005.

\bibitem[FK05]{FK05} Robert Fernholz and Ioannis Karatzas, {\it Relative arbitrage in volatility-stabilized markets}, Ann. Finance, 1:149–177, 2005.

\bibitem[FP14]{FP14} Andrea Frazzini and Lasse Heje Pedersen, {\it Betting against beta}, Journal of Financial Economics, 111(1):1–25, 2014.

\bibitem[KK20]{KK20} Ioannis Karatzas and Donghan Kim, {\it Trading strategies generated pathwise by functions of market weights}, Finance Stoch., 24:423–463, 2020.

\bibitem[KR17]{KR17} Ioannis Karatzas and Johannes Ruf, {\it Trading strategies generated by Lyapunov functions}, Finance Stoch., 21(3):753-787, 2017.

\bibitem[KS98]{KS98} Ioannis Karatzas and Steven Shreve, {\it Brownian Motion and Stochastic Calculus}, Springer, 1998.

\bibitem[KV15]{KV15} Ioannis Karatzas and Alexander Vervuurt, {\it Diversity-weighted portfolios with negative parameter}, Ann. Finance, 11:411–432, 2015.

\bibitem[L65]{L65} John Lintner, {\it Security prices, risk and maximal gains from diversification}, Journal of Finance, 20:587–615, 1965.

\bibitem[M52]{M52} Harry Markowitz, {\it Portfolio selection}, The Journal of Finance, 7(1):77–91, 1952.

\bibitem[M65]{M65} Jan Mossin, {\it Equilibrium in a capital asset market}, Econometrica, 35:768-783, 1965.

\bibitem[RX19]{RX19} Johannes Ruf and Kangjianan Xie, {\it Generalised Lyapunov functions and functionally generated trading strategies}, Appl. Math. Finance, 26(4):293-327, 2019.

\bibitem[RX20]{RX20} Johannes Ruf and Kangjianan Xie, {\it The impact of proportional transaction costs on systematically generated portfolios}, SIAM J. Financ. Math., 11(3):881-896, 2020.

\bibitem[S13]{S13} Winslow Strong, {\it Generalizations of Functionally Generated Portfolios with Applications to Statistical Arbitrage}, SIAM J. Financ. Math.,  5(1):472-492, 2014.

\bibitem[S64]{S64} William Sharpe, {\it Capital asset prices: A theory of market equilibrium under conditions of risk}, The Journal of Finance, 19(3):425–442, 1964.

\bibitem[SSV18]{SSV18} Alexander Schied, Leo Speiser and Iryna Voloshchenko, {\it Model-free portfolio theory and its functional master formula}, SIAM J. Financ. Math., 9(3):1074-1101, 2018.

\bibitem[V15]{V15} Alexander Vervuurt, {\it Topics in Stochastic Portfolio Theory}, arXiv:1504.02988, 2015.
\end{thebibliography}
\end{document}